\documentclass[12pt]{article}
\pdfoutput=1
\usepackage{jheppub}
\usepackage{amsmath,amssymb}
\usepackage{multirow}
\newcommand{\bea}{\begin{eqnarray}}
\newcommand{\eea}{\end{eqnarray}}
\newcommand{\Tr}{{\rm Tr}}

\begin{document}

\title{Hilbert Space of Finite $N$ Multi-matrix Models}

\author[a]{Robert de Mello Koch}
\affiliation[a]{School of Science, Huzhou University, Huzhou 313000, China}
\affiliation[a]{Mandelstam Institute for Theoretical Physics, School of Physics, University of the Witwatersrand, Private Bag 3, Wits 2050, South Africa}

\author[b]{and Antal Jevicki}
\affiliation[b]{Department of Physics, Brown University,
182 Hope Street, Providence, RI 02912, United States}
\affiliation[b]{Brown Theoretical Physics Center, Brown University,
340 Brook Street, Providence, RI 02912, United States}

\abstract{We study the Hilbert space structure of gauge-invariant operators emergent in large-$N$ multi-matrix quantum mechanics. Building on the framework of \cite{deMelloKoch:2025ngs}, we identify a class of light single-trace operators that behave like free creation operators at low energy but saturate beyond a critical excitation level, ceasing to generate new states. This $q$-reducibility is a direct consequence of finite N trace identities and leads to a dramatic truncation of the high-energy spectrum of the emergent theory. The resulting number of independent degrees of freedom is far smaller than naïve semiclassical expectations, providing a concrete mechanism for how nonperturbative constraints shape the ultraviolet behaviour of emergent theories.}

\maketitle
{\vskip -2.5cm}
\section{Introduction}

In \cite{deMelloKoch:2025ngs} we studied the algebra of $U(N)$ invariants in a general $d$-matrix model, with each matrix in the adjoint of the gauge group. The invariant single-trace (loop) operators are constrained at finite N by highly non-trivial trace identities. This is reflected in the partition function representing the Hilbert series of the ring of $U(N)$-invariant polynomials. For the multi-matrix oscillator, the partition function takes the universal form \cite{deMelloKoch:2025ngs}
\bea
Z(x)&=&\frac{1+\sum_i c_i x^i}{\prod_j (1-x^j)^{n_j}}\label{HironakaForm}
\eea
reflecting a Hironaka decomposition of the invariant ring: denominator factors correspond to primary invariants $P_A$ (which act freely and may be raised to any non-negative power), while numerator monomials correspond to secondary invariants $S_K$ (which appear only linearly) \cite{Sturmfels}. The Hochster–Roberts theorem \cite{HR} guarantees the existence of such a decomposition for rings of invariants under the action of linearly reductive groups, including $U(N)$. 
It follows that the Hilbert space of gauge-invariant operators admits the non-redundant decomposition
\bea
{\cal H}&=&\bigoplus_K\prod_A (P_A)^{n_A} S_K\label{LSstructure}
\eea
Although the partition function is derived from the oscillator, the structure in (\ref{LSstructure}) follows purely from trace relations and holds for multi-matrix quantum mechanics in general. Notice that the secondary invariants label the terms in the direct sum (\ref{LSstructure}) and that each such term is given by the free ring generated by the primary invariants. This suggests that the secondary invariants span the multiplicity space of representations of the free ring generated by the primaries.

The number of primary invariants equals the Krull dimension of the invariant ring. The number of secondary invariants—unknown in closed form—grows as $e^{c' N}$ in vector models \cite{deMelloKoch:2025cec} and as $e^{cN^2}$ in multi-matrix models \cite{deMelloKoch:2025qeq}, with $c',c$ order-one constants. For matrix models, this implies that most secondary invariants are composed from $O(N^2)$ fields. By analogy with the half-BPS sector of ${\cal N}=4$ SYM \cite{Lin:2004nb}, such heavy operators are expected to backreact and correspond to distinct semiclassical geometries, suggesting that secondary invariants encode nonperturbative gravitational degrees of freedom. We also anticipate secondary invariants composed of $O(N)$ fields, corresponding to operators dual to solitonic D-brane states such as giant gravitons\cite{McGreevy:2000cw}. This fits naturally with (\ref{LSstructure}): the primaries generate a Fock-space-like structure, while the secondaries invariants resemble nonperturbative states of the theory, reflecting solitons or new backgrounds.

In this paper we concentrate on Hilbert space aspects of this description. Specifically, at a large but finite $N$ there are secondary invariants constructed using $k$ fields with $k\ll N$. Such operators must be included among the generating invariants because, for single traces with fewer than $N$ matrices, the trace relations impose no constraints. These ``light" secondary invariants do not fit neatly into the solitonic or new background interpretation. This raises a question: What is the quantum mechanical meaning of these light secondary invariants, and how should we understand their role in the dual gravitational description?

It is natural to address   this question in the framework of  collective field theory, in which the  field variables are replaced by the complete set of invariants, initially treated as a free algebra. This description captures the large-$N$ dynamics exactly. At finite $N$, this is to be supplemented by  the trace relations, and fully solving these relations on the free invariant algebra yields the exact finite-$N$ Hilbert space. We carry out this program explicitly for $N$ identical bosons in $d$ spatial dimensions as a basic example, and then for two-matrix models at small $N$. In these tractable cases, the complete solution of the trace relations gives the full set of primary and secondary invariants. These concrete examples not only clarify the structure of the invariant ring but also motivate a definite physical interpretation for the light secondary invariants .

We focus on single trace secondary invariants constructed from a small ($\ll N$) number of fields. Powers of such small secondary operators also appear as higher-degree secondary invariants. However, there exists a maximal power beyond which additional products cease to produce independent invariants. This indicates that, while these secondaries initially behave like free Fock-space oscillators, they eventually saturate -- beyond a certain energy\footnote{We have not precisely characterized the saturation point. Clearly is must be beyond where there are $N$ matrices in the invariant, as saturation is a consequence of the trace relations. It is natural to expect saturation happens when the invariant is constructed using $O(N)$ fields.} they no longer generate new states. In the dual gravitational description, this implies that these degrees of freedom behave as perturbative gravitons in the infrared but decouple or become non-dynamical in the ultraviolet. The saturation mechanism imposes a substantial truncation of the high-energy spectrum, leaving far fewer degrees of freedom than naïve semiclassical gravity would predict.

This reduction is striking: the number of truly free generators is only $N^2+1$, compared to the exponentially large number of single-trace operators of length less than $N$. This sharp discrepancy highlights the central role of the trace relations in governing the ultraviolet behaviour of the theory.

To understand the mechanism behind the saturation phenomenon, we must revisit the role of trace relations. These relations imply that the secondary invariants are not algebraically independent but satisfy quadratic constraints of the form
\bea
S_I S_J=\sum_K f_K(\{P_A\})S_K
\eea
where $f_K(\{P_A\})$ are polynomials in the primary invariants and the identity is counted among the secondaries. In this sense, secondary invariants behave like emergent degrees of freedom with non-polynomial dependence on the primaries—analogous to solitonic excitations in field theory.

Crucially, this structure naturally produces saturation. Raising a small secondary invariant to a sufficiently high power $q$ yields an expression linear in the secondaries and polynomial in the primaries. Beyond this threshold, repeated applications of the same small secondary reproduce states already present, and must be excluded to maintain a non-redundant Hilbert space basis.

This motivates interpreting primitive secondary invariants as \emph{q-reducible oscillators}: they act like independent creation operators at low excitation levels, but beyond a critical power their action becomes redundant. Their finite range of action distinguishes them sharply from the genuinely free (Fock-space) degrees of freedom generated by the primaries.

The paper is organized as follows: In Section \ref{HS} we introduce the collective field framework, paying special attention to how the trace relations are realized as constraints that commute with the Hamiltonian. The finite $N$ Hilbert space follows by imposing the constraints on the collective fields, which are given by the free algebra of invariants. In Section \ref{bosons} we analyze invariant rings associated with $N$ identical bosons in $d$ spatial dimensions. We examine three progressively more intricate examples, each of which exhibits the phenomenon of $q$-reducible oscillators. These models are sufficiently simple to permit a relatively transparent analysis. In Section \ref{MatrixModels} we turn to matrix models, which present significantly more complexity. Our first example involves the invariant ring of two $3\times 3$ matrices, which possesses 10 primary invariants and a non-trivial single secondary invariant. We demonstrate how this secondary invariant gives rise to a reducible oscillator. To explore a more representative and richer structure, we then study the invariants of two $4\times 4$ matrices. In this example, the full ring of gauge-invariant operators is generated by 17 primary invariants and 48 secondary invariants. Fortunately, the explicit construction of these invariants was recently achieved by mathematicians using techniques from non-commutative Poisson geometry \cite{pgeometry}. This more elaborate model reveals multiple instances of $q$-reducible oscillators, providing a fertile ground for understanding the truncation mechanism in operator algebras. Finally, we give some discussion of our results in Section \ref{discussion}.

\section{Hilbert Space}\label{HS}

Collective field theory \cite{JS1} provides a field theory representation of large $N$ multi-matrix $X^a(t)$ $a=1,2,\cdots,d$ systems. The canonical picture of collective field theory provides equal time evolution. It is based on equal time loops ($t=0$) and their canonical conjugates. These are given as
\bea
\phi(c)&=&\Tr(X^{a_1}X^{a_2}X^{a_3}\cdots)
\eea
where $c$ is any given word constructed using the $X^a$ as letters. Due to cyclicity of the trace, words related by a cyclic permutation must be identified. We refer to the number of matrices in the trace as the length $l(c)$ of $\phi(c)$. The conjugate is simply
\bea
\bar{\phi}(c)&\equiv&{\partial\over\partial\phi(c)}\qquad [\bar{\phi}(c),\phi(c')]\,\,=\,\,\delta(c,c')
\eea
These play the role of creation/annihilation operators in the loop space description. The collective Hamiltonian then takes the form
\bea
\hat{H}&=&N\sum_c l(c) \phi(c)\bar{\phi}(c)\,+\,\sum_{c,c_1.c_2} J(c;c_1,c_2)\phi(c)\bar{\phi}(c_1)\bar{\phi}(c_2)\,+\,V
\eea
where $V$ allows for a general interaction.This formulation of the large N theory allows for analytical and numerical evaluation of 
energy levels and the spectrum\cite{Koch:2021yeb}. A thermofield extension was also given\cite{Jevicki:2025ybt}.

Even though the theory is based on an infinite set of all single trace 
operators there is a sense in which it continues to be valid at specific, finite $N$. Namely, we have $1/N$ as a coupling constant  in
the Hamiltonian and the finite N trace relations are respected.

For finite integer $N$, they show up as an infinite sequence of constraints commuting with the collective Hamiltonian
\bea
[\Gamma_\alpha(\{\phi\}),\hat{H}]\,\,=\,\,[\Gamma_\alpha(\{\phi\}),H_2+H_3]&=&0\label{ConcComm}
\eea
independently of the specific interactions. These constraints arise from trace relations -- identities satisfied by any $N\times N$ matrix -- which depend only on $N$ and are entirely interaction-independent. In the overcomplete collective field theory \cite{JS1} these relations are not imposed. Here we impose them to determine the structure of the finite-$N$ Hilbert space. Because our analysis depends solely on the trace relations, the results apply universally to a wide class of potentials, from the trace of a commutator squared, to double-trace terms and potentials that are general polynomial functions of the $X^a$.

That the collective Hamiltonian always commutes with the finite-$N$ trace relations reflects a deep result of Procesi \cite{Procesi2}. The overcomplete collective field theory describes dynamics on the space of invariant fields \cite{JS1} without imposing any trace relations—that is, it is formulated at the level of the free trace algebra. Procesi’s result which supplies a formal inverse to the Cayley–Hamilton theorem \cite{Procesi2}, makes this structure explicit: it implies that the algebra of $U(N)$-invariant polynomial functions (which is the finite $N$ Hilbert space we are constructing) is obtained by starting from the free trace algebra (which is the space of collective fields of the over complete collective field theory) and imposing all trace relations. Thus, at finite $N$, trace relations are inescapable, expressing universal\footnote{A PI algebra (Polynomial Identity algebra) is a ring or algebra satisfying a nontrivial polynomial identity in noncommuting variables.} PI-constraints rather than model-dependent dynamics.

A direct way to understand these constraints is through the fact that $H_{\rm coll}=H_2+H_3$ is exactly solvable. The growing (polynomial) set of eigenstates can be constructed using degenerate perturbation theory \cite{Jevicki:1991yi}. This provides a basis known as the Schur basis \cite{Jevicki:1991yi,Corley:2001zk} for a single matrix or restricted Schur basis \cite{Bhattacharyya:2008rb} for multi matrix models\footnote{For other bases that are also labelled by Young diagrams and so may play a similar role see for example \cite{Brown:2007xh}.}. For fixed (integer) $N$ the subsidiary conditions in (\ref{ConcComm}) are states that become null when $N$ is integer. They are also in one-to-one relation with finite $N$ trace relations. Through a comprehensive analysis of these finite $N$ relations the following \emph{decomposition} was established in \cite{deMelloKoch:2025ngs}:
\begin{itemize}
\item[1.] Loops $\phi(c)$ which are unconstrained. We denote them as $\varphi(c)$ and call them \emph{gravitons}.
\item[2.] Loop variables $\phi(c'')$ which obey quadratic constraints as introduced in \cite{deMelloKoch:2025ngs}. We call these \emph{solitons}. They obey relations of the form\footnote{If we include the identity among the $\psi(c'')$, then there is no need for $P_1(\varphi)$ in (\ref{qdrel}).}
\bea
\psi(c_1'')\psi(c_2'')&=&\sum_{c''}\psi(c'')P_{c''}(\varphi)+P_1(\varphi)\label{qdrel}
\eea  
\end{itemize}
The Fock space them decomposes into $N_s=0$ and $1$ (soliton) sectors
\bea
\prod_i\hat{\varphi}(c_i)|0\rangle\qquad N_s\,\,=\,\,0
\eea
\bea
\left(\prod_i\hat{\varphi}(c_i)\right)\hat{\psi}(c_s)|0\rangle\qquad N_s\,\,=\,\,1
\eea
Due to the quadratic relations only one soliton operator is allowed. This will have immediate consequence on the thermodynamics of this collective system, with the property that the states created by $\hat{\psi}$'s survive but multi-soliton states are forbidden. The partition function then naturally takes the form
\bea
Z(q)={N(q)\over D(q)}
\eea
with
\bea
N(q)&=&\sum_s n(s) q^{l(s)}
\eea
and where the growth in the number of solitons is
\bea
N(1)&=&e^{c N^2}
\eea
with $c$ an order 1 number.

 The unconstrained collective field theory, realized at $N=\infty$, corresponds to the stable sector governed by the primary invariants. At finite $N$ the stable sector receives contributions from both primary and secondary invariants. This connection is clearly illustrated in \cite{Djokovic}, where the free energy of the large $N$ free loop gas is recovered from the stable part of the Molien–Weyl partition function. Returning to the finite $N$ structure of the collective Hilbert space, we now have a finite system of standard oscillators (corresponding to primary traces), supplemented by a larger number of q-reducible oscillators (given by the generating set of secondary invariants). The finite nature of primary oscillators and the q-reducibility of secondary ones implies a finiteness properties of this field theory, in particular the entanglement entropy.
Regarding q-deformation we recall that such properties were encountered  in the invariant (finite $N$) description of $S_N$ orbifold\cite {Jevicki:1998rr} field theories. In that case this property was argued to
result in non-commutative emergent space-time.

\section{$S_N$ invariance for $N$ bosons in $d$ dimensions}\label{bosons}

Consider a system of $N$ identical bosons in $d$ spatial dimensions. The configuration of the system is described by $N$ independent $d$-dimensional vectors $\vec{x}_i=(x^{1}_i, x^{2}_i,\cdots, x^{d}_i)$, where $i=1,\cdots,N$. These vectors represent the positions of the bosons, with each index $i$ labeling a particle. Requiring invariance under the symmetric group $S_N$, which acts on the particle labels $i$, captures the indistinguishability of the bosons.

Altogether, the system is described by $Nd$ variables. Let $\mathbb{C}[V^{Nd}]$ denote the ring of complex polynomials in these $Nd$ variables. We are interested in the subring of polynomials invariant under the action of $S_ N$, denoted by $\mathbb{C}[V^{Nd}]^{S_N}$. This subring is known as the ring of multisymmetric polynomials.

In Section~\ref{collective}, we outline the collective field theory formulation for this bosonic system. This overcomplete description is defined on the full space of invariant (collective) fields. To restrict to a fixed, finite number $N$ of bosons, one must impose specific constraints on this space. We give a complete description of the required constraints. In the remainder of this section, we solve these constraints explicitly in several illustrative examples. It is through this enforcement that $q$-reducible oscillators naturally appear. Furthermore, we show that the space of $S_N$-invariant operators is generated by primary and secondary invariants, a structure derived solely from the trace relations. As a result, this organization is entirely general and independent of the particular interactions governing the bosons.

\subsection{Collective Description}\label{collective}

We consider a system described by the $N$-body Hamiltonian 
\bea
H&=&{1\over 2}\sum_{i=1}^N\sum_{a=1}^d \Big(-{\partial\over\partial x_i^a}{\partial\over\partial x_i^a}+\omega^2
(x_i^a)^2\Big)+V(x_i^a)\label{bosonhamiltonian}
\eea
where $V(x_i^a)$ is any potential invariant under the action of $S_N$ on the particle label $i$. The collective fields are given by the complete set of $S_N$-invariant combinations of the coordinates. For this example, a convenient choice is
\bea
\phi(n_1,n_2,\cdots,n_d)&=&\sum_{i=1}^N (x_i^1)^{n_1}
(x_i^2)^{n_2}\cdots (x_i^d)^{n_d}
\eea
For any fixed $N$ these invariants are over complete. To derive the relations between them, introduce the $d$ $N\times N$ matrices defined by
\bea
X^a&\equiv&\left[\begin{array}{ccccc}
     x^a_1&0 &0  &\cdots &0\\
     0    &x^a_2 &0 &\cdots &0\\
     0    &0     &x^a_3 &\cdots &0\\
     \vdots &\vdots &\vdots &\ddots &0\\
     0 &0 &0 &\cdots &x^a_N
\end{array}\right]\label{matsdefined}
\eea
The relations between the invariant variables now follow by anti symmetrizing the column indices in the expression
\bea
\sum_{\sigma\in S_N}{\rm sgn}(\sigma)(W_1)_{i_1i_{\sigma(1)}}(W_2)_{i_2 i_{\sigma(2)}}\cdots (W_{N+1})_{i_{N+1}i_{\sigma(N+1)}}=0\label{PermCayleyHamilton}
\eea
where ${\rm sgn}(\sigma)$ is the parity of $\sigma$, and where the $W_a$ are each any word constructed out of the $X^a$. This identity is true because antisymmetrizing $N+1$ indices that each take $N$ values always vanishes. Thus the relation (\ref{PermCayleyHamilton}) holds irresptive of the choice of $V(x_i^a)$ in (\ref{bosonhamiltonian}). The main goal of this paper is to completely solve the full set of relations (\ref{PermCayleyHamilton}), thereby determining the structure of the finite-$N$ Hilbert space.

\subsection{$d=2$ and $N=2$}

In this section, we examine the simplest example in which $q$-reducible oscillators arise. We will denote $x^1_i$ by $x_i$ and $x^2_i$ by $y_i$. If we grade by $x$ and $y$, a simple application of the Molien-Weyl formula leads to the following partition function
\bea
Z(t_1,t_2)=\frac{1+t_{1} t_{2}}{(1-t_{1}) \left(1-t_{1}^2\right) (1-t_{2}) \left(1-t_{2}^2\right)}\label{PFd2N2}
\eea
This Hilbert series suggests that the algebra of invariants $\mathbb{C}[V^{Nd}]^{S_N}$ admits a Hironaka decomposition, constructed using 4 primary invariants and a single non-trivial secondary invariant. The graded partition function suggests the following primary invariants
\bea
P_1&=&x_1+x_2\qquad P_2\,\,=\,\,(x_1)^2+(x_2)^2\cr\cr P_3&=&y_1+y_2\qquad P_4\,\,=\,\,(y_1)^2+(y_2)^2
\eea
while the secondary invariants are given by
\bea
S_0&=&1\qquad S_1\,\,=\,\,x_1 y_1+x_2 y_2
\eea
To establish this ansatz, we must use the trace relations to show that these generate the complete space of invariants. Note that for $N=2$ the trace relations take the following form
\bea
T_2(A,B,C)&=&\Tr(A)\Tr(B)\Tr(C)-\Tr(AB)\Tr(C)-\Tr(A)\Tr(BC)\cr\cr
&&-\Tr(B)\Tr(AC)+\Tr(ABC)+\Tr(ACB)\label{tracerelationN2}
\eea
where $A,B$ and $C$ are any words in $X$ and $Y$, with $X,Y$ defined as in (\ref{matsdefined}). The complete set of degree 0 invariants is given by $S_0$. The complete space of degree 1 invariants are given by $P_1$ and $P_2$. The complete space of degree 2 invariants are given by $P_3$, $P_4$ and $S_2$. The trace relation $T_2(X,X,X)=0$ implies that
\bea
x_1^3+x_2^3&=&{3\over 2}P_1P_2={1\over 2}P_1^3
\eea
Similarly, $T_2(Y,Y,Y)=0$ implies that
\bea
y_1^3+y_2^3&=&{3\over 2}P_3P_4={1\over 2}P_3^3
\eea
Next, the trace relation $T_2(X,X,Y)=0$ implies that
\bea
x_1^2y_1+x_2^2y_2&=&P_1 S_1+{1\over 2}P_2P_3-\frac12 P_1^2 P_3
\eea
while $T_2(X,Y,Y)=0$ implies that
\bea
x_1y_1^2+x_2y_2^2&=&P_3 S_1+{1\over 2}P_4P_1-\frac12 P_3^2 P_1
\eea
This completes our description of the degree-3 invariants in terms of the chosen generating set. Extending the analysis to degree 4, degree 5, and higher is straightforward, but not essential. The partition function (\ref{PFd2N2}) already specifies exactly how many invariants exist at each degree. Once the generators have been verified at low degree, the Hironaka decomposition guarantees that they will generate the complete space of invariants at all higher degrees.

$S_1$ is the simplest example of a reducible oscillator. Since $S_1$ appears in the generating set, as a secondary invariant, it can occur only linearly in the expression for any invariant. The partition function (\ref{PFd2N2}) can be rearranged to make this q-reducible oscillator more explicit, i.e. so that $S_1$ appears associated to a factor in the denominator rather than the numerator. This is achieved by multiplying the partition function by the identity
\bea
Z(t_1,t_2)=\frac{1-(t_1 t_2)^2}{(1-t_{1}) \left(1-t_{1}^2\right) (1-t_{2}) \left(1-t_{2}^2\right)(1-t_1 t_2)}
\eea
Notice that there are now terms in the denominator for all four primary invariants, as well as for $S_1$. The appearance of a negative term in the numerator signals a constraint, which is of degree four and is written as
\bea
(S_1)^2&=&S_1 P_1 P_3+S_0\left(P_2 P_4-\frac{1}{2}P_1^2 P_4-\frac{1}{2} P_2 P_3^2\right)
\eea
This constraint is again a consequence of the trace relations. It can be solved to determine $S_1$ in terms of the primary invariants. There are two solutions for $S_1$ given by
\bea
(S_1)_{\pm}&=&\frac12 P_1 P_3\pm\frac12\sqrt{P_1^2P_3^2+4P_2P_4-2P_1^2P_4-2P_2P_3^2}
\eea
Either solution can be chosen because the sum of the two solutions is a polynomial in the primary invariants
\bea
(S_1)_++(S_1)_-=P_1P_3
\eea

Squaring the non-trivial secondary invariant already leads to a redundancy. This is special to this simple example. In general, we will see that $q$-reducible oscillators with $q>1$ are common.

\subsection{$N=3$ and $d=2$}\label{NiceQRep}

We now consider an example which is complicated enough to exhibit the full structure of the problem. The Hilbert series, again graded by $x$ and $y$, is given by
\bea
Z(t_1,t_2)&=&\frac{1+t_1t_2+t_1^2 t_2+t_1t_2^2+t_1^2 t_2^2+t_1^3 t_2^3}{(1-t_1)(1-t_1^2)(1-t_1^3)(1-t_2)(1-t_2^2)(1-t_2^3)}\label{OrgPart}
\eea
The Hironaka decomposition uses 6 primary invariants and 5 non-trivial secondary invariants. From the above Hilbert series, we know the grading of each generator. Every invariant of degree 3 or less must be included amongst the generating set because the trace relations only start at degree 4. Thus, we immediately find the following formulas for the primary invariants
\bea
P_1&=&x_1+x_2+x_3\quad 
P_2\,\,=\,\,(x_1)^2+(x_2)^2+(x_3)^2\quad 
P_3\,\,=\,\,(x_1)^3+(x_2)^3+(x_3)^3\cr\cr
P_4&=&y_1+y_2+y_3\quad 
P_5=(y_1)^2+(y_2)^2+(y_3)^2\quad
P_6=(y_1)^3+(y_2)^3+(y_3)^3\label{PIsN3D2}
\eea
the following formulas for the secondary invariants
\bea
S_0&=&1\qquad S_1=x_1y_1+x_2y_2+x_3y_3\qquad S_2\,\,=\,\,(x_1)^2 y_1+(x_2)^2y_2+(x_3)^2y_3\cr\cr
S_3&=&x_1(y_1)^2+x_2(y_2)^2+x_3(y_3)^2
\eea
There are two secondary invariants that are not yet determined. As always, these are determined using the trace relations, which are given by setting $N=3$ in (\ref{PermCayleyHamilton}) to obtain
\bea
0&=&T_3(W_1,W_2,W_3,W_4)\cr\cr
&=&\Tr(W_1)\Tr(W_2)\Tr(W_3)\Tr(W_4)
-\Tr(W_1 W_2)\Tr(W_3)\Tr(W_4)-\Tr(W_1 W_3)\Tr(W_2)\Tr(W_4)\cr\cr
&-&\Tr(W_1 W_4)\Tr(W_2)\Tr(W_3)
-\Tr(W_2 W_3)\Tr(W_1)\Tr(W_4)
-\Tr(W_2 W_4)\Tr(W_1)\Tr(W_3)\cr\cr
&-&\Tr(W_3 W_4)\Tr(W_1)\Tr(W_2)
+\Tr(W_1 W_2 W_3)\Tr(W_4)+\Tr(W_1 W_3 W_2)\Tr(W_4)\cr\cr
&+&\Tr(W_1 W_2 W_4)\Tr(W_3)+\Tr(W_1 W_4 W_2) \Tr(W_3)+\Tr(W_1 W_3 W_4)\Tr(W_2)\cr\cr
&+&\Tr(W_1 W_4 W_3) \Tr(W_2)+\Tr(W_2 W_3 W_4)\Tr(W_1)+\Tr(W_2 W_4 W_3)\Tr(W_1)\cr\cr
&+&\Tr(W_1 W_2)\Tr(W_3W_4)+\Tr(W_1 W_3)\Tr(W_2W_4)+\Tr(W_1 W_4)\Tr(W_2W_3)\cr\cr
&-&\Tr(W_1 W_2 W_3W_4)-\Tr(W_1 W_2 W_4W_3)-\Tr(W_1 W_3 W_2W_4)-\Tr(W_1 W_3 W_4 W_2)\cr\cr
&-&\Tr(W_1 W_4 W_2 W_3)-\Tr(W_1 W_4 W_3 W_2)
\eea
The complete details of how this trace relation is used to determine the complete set of invariants is described in Appendix \ref{DeriveGens}. Here we summarize some of the key features of this derivation.

The trace relations are first applicable at degree 4. At this degree, we find that the trace relation $T_3(X,X,Y,Y)$ only determines the invariant
\bea
(x_1)^2(y_1)^2+(x_2)^2(y_2)^2+(x_2)^2(y_2)^2
\eea
if we introduce an additional secondary invariant
\bea
S_4&=&S_1^2
\eea
With the help of this additional secondary invariant, the complete set of invariants at degree 4 can be generated. At degree 5 we find that all invariants can be generated using our generating set. At degree 6 we again find that we need to introduce one more secondary invariant, given by
\bea
 S_5&=&S_1^3
\eea
Notice that $S_4$ and $S_5$ are both multi particle secondary invariants i.e. they can be expressed as a product of lower degree secondary invariants. For more details, see Appendix \ref{DeriveGens}. This Appendix verifies that (i) any expression quadratic in the secondary invariants can be expressed as a sum of terms linear in the secondary invariants and (ii) trace relations allow us to write all trace invariants in terms of the above generators as dictated by the Hironaka decomposition. In checking that all $S_N$ invariant operators can be written in terms of the set of invariants, new invariants are introduced as they are needed.

To rephrase the description of the algebra of invariants in terms of the primary and and a smaller set of secondary invariants, multiply by
\bea
1=\frac{(1-t_1t_2) (1-t_1^2t_2) (1-t_1t_2^2)}{(1-t_1t_2) (1-t_1^2t_2) (1-t_1t_2^2)}
\eea
which leads to the following partition function
\bea
Z(t_1,t_2)&=&\frac{N(t_1,t_2)}{(1-t_1)(1-t_1^2)(1-t_1^3)(1-t_2)(1-t_2^2)(1-t_2^3)(1-t_1t_2) (1-t_1^2t_2) (1-t_1t_2^2)}\cr\cr
&&
\eea
with
\bea
N(t_1,t_2)&=&1-t_1^3t_2^2-t_1^4 t_2^2-t_1^2t_2^3-t_1^2t_2^4-t_1^3t_2^3-t_1^4t_2^4\cr\cr &&+t_1^5t_2^3+2t_1^4t_2^4+t_1^5t_2^4+ t_1^3t_2^5+t_1^4t_2^5+t_1^6t_2^5+t_1^5t_2^6\cr\cr
&&-t_1^5t_2^6-t_1^6t_2^5-t_1^7t_2^7
\eea
Notice that all primaries together with $S_1$, $S_2$ and $S_3$ are associated to factors in the denominator. The multi particle secondary invariants $S_4$ and $S_5$ are not included separately but rather are generated by taking powers of $S_1$. The complexity of the numerator in the Hilbert series reflects the presence of non-trivial relations among the generators. In Appendix \ref{confirm} we have confirmed that the numerator does indeed correctly encode the constraints associated with the primary and single particle secondary invariants. In summary, the oscillators of the $N=3$ collective field theory are the primary invariants (\ref{PIsN3D2}), which can be raised to any power, together with $S_1,(S_1)^2,(S_1)^3,S_2$ and $S_3$. This completes the construction of the finite $N$ Hilbert space.

Looking at the numerator of the original partition function (\ref{OrgPart}), the contribution of the secondary invariants to the collective field theory oscillators are already visible
\bea
1+t_1 t_2+t_1^2 t_2^2+t_1^3 t_2^3+t_1^2 t_2+t_1t_2^2\leftrightarrow
S_{0}+S_{1}+(S_{1})^2+(S_{1})^3+S_{2}+S_{3}\cr
\eea
where the grading picks out a unique association between terms on the left and right hand sides. As $N$ is increased the number of $q$-reducible oscillators grows much more rapidly than the growth in the number of the usual oscillators and for reasonably large values of $N$ most oscillators are $q$-reducible with only a small fraction corresponding to free oscillators and represented by primary invariants.

Finally, note that there is some freedom in choosing the oscillators. For example, we could have chosen to eliminate $(S_{1}^\dagger)^3$, as long as we keep $S_{2}^\dagger S_{3}^\dagger$.

\subsection{$N=3$ and $d=3$}

We denote $x^1_i$ by $x_i$, $x^2_i$ by $y_i$ and $x^3_i$ by $z_i$ and grade on all three letters. The graded Hilbert series is given by
\bea
Z(t_1,t_2,t_3)&=&\frac{N(t_1,t_2,t_3)}{(1-t_{1}) \left(1-t_{1}^2\right) \left(1-t_{1}^3\right) (1-t_{2}) \left(1-t_{2}^2\right) \left(1-t_{2}^3\right) (1-t_{3}) \left(1-t_{3}^2\right) \left(1-t_{3}^3\right)}\cr\cr
&&
\eea
where
\bea
N(t_1,t_2,t_3)&=&1+t_1 t_2+t_1^2 t_2+t_1 t_2^2+t_1^2 t_2^2+t_1^3 t_2^3+t_1 t_3+t_1^2 t_3+t_2 t_3+t_1 t_2 t_3+t_1^2 t_2 t_3+t_1^3 t_2 t_3\cr\cr
&&+t_2^2 t_3+t_1 t_2^2 t_3+t_1^2 t_2^2 t_3+t_1^3 t_2^2 t_3+t_1 t_2^3 t_3+t_1^2 t_2^3 t_3+t_1 t_3^2+t_1^2 t_3^2+t_2 t_3^2+t_1 t_2 t_3^2\cr\cr
&&+t_1^2 t_2 t_3^2+t_1^3 t_2 t_3^2+t_2^2 t_3^2+t_1 t_2^2 t_3^2+t_1^2 t_2^2 t_3^2+t_1^3 t_2^2 t_3^2+t_1 t_2^3 t_3^2+t_1^2 t_2^3 t_3^2+t_1^3 t_3^3+t_1 t_2 t_3^3\cr\cr
&&+t_1^2 t_2 t_3^3+t_1 t_2^2 t_3^3+t_1^2 t_2^2 t_3^3+t_2^3 t_3^3
\eea
The Hironaka decomposition uses 9 primary invariants and 35 non-trivial secondary invariants. Given the lesson of the previous subsection, to obtain the physical presentation of these invariants we simply interpret the numerator. Each term in the numerator (shown below) has a unique interpretation, thanks to the grading.
\bea
\{t_1 t_2,t_1^2 t_2^2,t_1^3 t_2^3\}&\leftrightarrow&\{S_{1},(S_{1})^2,(S_{1})^3\}\cr\cr
\{t_1 t_3,t_1^2 t_3^2,t_1^3 t_3^3\}&\leftrightarrow&\{S_{2},(S_{2})^2,(S_{2})^3\}\cr\cr
\{t_2 t_3,t_2^2 t_3^2,t_2^3 t_3^3\}&\leftrightarrow&\{S_{3},(S_{3})^2,(S_{3})^3\}\cr\cr
\{t_1^2 t_2,t_1 t_2^2\}&\leftrightarrow&\{S_{4},S_{5}\}\cr\cr
\{t_1^2 t_3,t_1 t_3^2\}&\leftrightarrow&\{S_{6},S_{7}\}\cr\cr
\{t_2^2 t_3,t_2 t_3^2\}&\leftrightarrow&\{S_{8},S_{9}\}\cr\cr
\{t_1 t_2 t_3,t_1^2 t_2^2 t_3^2\}&\leftrightarrow&\{S_{10}\}\cr\cr
\{t_1^2 t_2 t_3,t_1 t_2^2 t_3,t_1 t_2 t_3^2\}&\leftrightarrow&\{S_{11},S_{12},S_{13}\}\cr\cr
\{t_1^3 t_2 t_3,t_1 t_2^3 t_3,t_1 t_2 t_3^3\}&\leftrightarrow&\{S_{14},S_{15},S_{16}\}\cr\cr
\{t_1^2 t_2^2 t_3,t_1^2 t_2 t_3^2,t_1 t_2^2 t_3^2\}&\leftrightarrow&\{S_{17},S_{18},S_{19}\}\cr\cr
\{t_1^3 t_2^2 t_3,t_1^2 t_2^3 t_3,t_1^3 t_2 t_3^2,t_1 t_2^3 t_3^2,t_1^2 t_2 t_3^3,t_1 t_2^2 t_3^3\}&\leftrightarrow&\{S_{20},S_{21},S_{22},S_{23},S_{24},S_{25}\}\cr\cr
\{t_1^3 t_2^2 t_3^2,t_1^2 t_2^3 t_3^2,t_1^2 t_2^2 t_3^3\}&\leftrightarrow&\{S_{26},S_{27},S_{28}\}
\eea
where
\bea
S_{1}&=&\sum_{i=1}^3 x_i y_i\qquad
S_{2}\,\,=\,\,\sum_{i=1}^3 x_i z_i\qquad
S_{3}\,\,=\,\,\sum_{i=1}^3 y_i z_i\cr\cr
S_{4}&=&\sum_{i=1}^3 (x_i)^2 y_i\qquad
S_{5}\,\,=\,\,\sum_{i=1}^3 (y_i)^2 x_i\qquad\cr\cr
S_{6}&=&\sum_{i=1}^3 (x_i)^2 z_i\qquad
S_{7}\,\,=\,\,\sum_{i=1}^3 (z_i)^2 x_i\qquad\cr\cr
S_{8}&=&\sum_{i=1}^3 (y_i)^2 z_i\qquad
S_{9}\,\,=\,\,\sum_{i=1}^3 (z_i)^2 y_i\qquad\cr\cr
S_{10}&=&\sum_{i=1}^3 x_iy_i z_i\cr\cr
S_{11}&=&\sum_{i=1}^3 (x_i)^2 y_i z_i\qquad
S_{12}\,\,=\,\,\sum_{i=1}^3 x_i(y_i)^2 z_i\qquad
S_{13}\,\,=\,\,\sum_{i=1}^3 x_iy_i (z_i)^2\cr\cr
S_{14}&=&\sum_{i=1}^3 (x_i)^3 y_i z_i\qquad
S_{15}\,\,=\,\,\sum_{i=1}^3 x_i(y_i)^3 z_i\qquad
S_{16}\,\,=\,\,\sum_{i=1}^3 x_iy_i (z_i)^3\cr\cr
S_{17}&=&\sum_{i=1}^3 (x_i)^2 (y_i)^2 z_i\qquad
S_{18}\,\,=\,\,\sum_{i=1}^3 (x_i)^2 y_i (z_i)^2\qquad
S_{19}\,\,=\,\,\sum_{i=1}^3 x_i(y_i)^2 (z_i)^2\cr\cr
S_{17}&=&\sum_{i=1}^3 (x_i)^2 (y_i)^2 z_i\qquad
S_{18}\,\,=\,\,\sum_{i=1}^3 (x_i)^2 y_i (z_i)^2\qquad
S_{19}\,\,=\,\,\sum_{i=1}^3 x_i(y_i)^2 (z_i)^2\cr\cr
S_{20}&=&\sum_{i=1}^3 (x_i)^3 (y_i)^2 z_i\qquad
S_{21}\,\,=\,\,\sum_{i=1}^3 (x_i)^2 (y_i)^3 z_i\qquad
S_{22}\,\,=\,\,\sum_{i=1}^3 (x_i)^3 y_i (z_i)^2\cr\cr
S_{23}&=&\sum_{i=1}^3 x_i (y_i)^3 (z_i)^2\qquad
S_{24}\,\,=\,\,\sum_{i=1}^3 (x_i)^2 y_i (z_i)^3\qquad
S_{25}\,\,=\,\,\sum_{i=1}^3 x_i(y_i)^2 (z_i)^3\cr\cr
S_{26}&=&\sum_{i=1}^3 (x_i)^3 (y_i)^2 (z_i)^2\quad
S_{27}\,=\,\sum_{i=1}^3 (x_i)^2 (y_i)^3 (z_i)^2\quad
S_{28}\,=\,\sum_{i=1}^3 (x_i)^2(y_i)^2 (z_i)^3\cr\cr
&&
\eea
One could, once again, prove this proposal by using the trace relations to show that these oscillators -- along with their associated saturation bounds -- generate the complete space of gauge-invariant operators. Further, it is possible to express the partition function using these $q$-reducible oscillators as generators. In such a rewriting, the partition function acquires a nontrivial numerator, reflecting the intricate structure of the constraints among the generators. This expression can then be explicitly verified by analyzing the algebraic relations satisfied by the generators, as well as the higher-order relations among those constraints.
To recapitulate ,in terms of collective field theory:we have that it consist of 9 primary oscillators $P_i$ and 28 q-oscillators $S_i$ .On the Fock space the primary oscillators are freely  acting while the
secondary q-oscillators act as summarized in Eq.3.30 .

The discussion of the examples considered in this section makes it clear that we can trade multiple secondary invariants for a single $q$-reducible oscillator. As $N$ is increased $q$ increases, i.e. these operators can be raised to higher powers before they are saturated. When $N=\infty$ they become ordinary oscillators. Oscillators that can be raised to a very high power before they saturate are behaving like perturbative modes. Oscillators that saturate after they are raised to a small power are behaving more like non-perturbative states.

\section{Matrix Models}\label{MatrixModels}

In this section we will consider the algebra of $U(N)$ invariants for a matrix model of two $N\times N$ matrices. This algebra is usually denoted by $C_{Nd}$. For simplicity we again consider a matrix oscillator, denoting the two matrices as $X$ and $Y$. Choosing $N=2$ is too simple to be interesting as the resulting algebra has no non-trivial secondary invariants. For $N=3$ there are 10 primary invariants and 2 secondary invariants one of which is non-trivial. This model, which is discussed in Section \ref{Nis3}, does not illustrate the general case. In Section \ref{Nis4} we consider the model with $N=4$. This model is considerably more complicated and the algebra of invariants is generated by 17 primary invariants and 48 secondary invariants. Fortunately the explicit form of this complete set of generators has recently been computed in \cite{pgeometry}. We will be able to make use of these results to exhibit a spectrum of $q$ reducible oscillators. For $N\ge 5$ the model has thousands of secondary invariants, so it is simply not practical to repeat the analysis.

\subsection{$N=3$}\label{Nis3}

This model has been studied in detail in \cite{T}. The Molien-Weyl partition function for 2 $3\times 3$ matrices is given by
\bea
Z(x) &=& \frac{1 + x^6}{(1-x)^2(1-x^2)^3(1-x^3)^4(1-x^4)}\label{N3M2}
\eea
This makes it clear that the Hironaka decomposition uses 10 primary invariants and two secondary invariants. We know that all traces with $\le N=3$ matrices must be included in the trace. Consequently, we immediately find the following 9 primary invariants
\bea
P_1&=&\Tr (X)\qquad P_2\,\,=\,\,\Tr (Y)\cr\cr
P_3&=&\Tr (X^2)\qquad P_4\,\,=\,\,\Tr (XY)\qquad P_5\,\,=\,\,\Tr (Y^2)\cr\cr
P_6 &=& \Tr(X^3), \qquad P_7 = \Tr(X^2 Y), \qquad P_8 = \Tr(XY^2),\qquad 
P_9 \,\,=\,\, \Tr(Y^3)
\eea
and the trivial secondary invariant
\bea
S_0 &=&1
\label{solitonN3M2}
\eea

To proceed, we use the trace relations to determine the gauge invariant operators, in terms of the generating set, adding new generators as required until we have accounted for all invariants appearing in the Hilbert series. At degree 4 the relevant single trace operators are constructed from $n$ $X$s and $m$ $Y$s. In the table below we summarize the trace relations at each $(n,m)$ as well as the possible single trace operators. For $n=2=m$ there are two independent single trace operators that can be defined. Since we only have a single trace relation, one of these (or a linear comination of them) must be added as an invariant. We choose to add
\bea
P_{10}=\Tr(X^2Y^2)
\eea
as an extra invariant. Looking at the formula (\ref{N3M2}) we know that this is a primary invariant.

\begin{table}[h]
\centering
\begin{tabular}{|c|c|c|c|}
\hline
$(n,m)$ &Independent Trace Relations & Single Trace Operators & New Invariants \\ \hline
$(4,0)$ &$T_3(X,X,X,X)=0$ & $\Tr(X^4)$ & none \\ \hline
$(3,1)$ &$T_3(X,X,X,Y)=0$ & $\Tr(X^3Y)$ & none \\ \hline
$(2,2)$ &$T_3(X,X,Y,Y)=0$ & $\Tr(X^2Y^2)$, $\Tr(XYXY)$ & $\Tr(X^2Y^2)$ \\ \hline
$(1,3)$ &$T_3(X,Y,Y,Y)=0$ & $\Tr(XY^3)$ & none \\ \hline
$(0,4)$ &$T_3(Y,Y,Y,Y)=0$ & $\Tr(Y^4)$ & none \\ \hline
\end{tabular}
\caption{There is a single trace relation for invariants constructed $n$ $X$s and $m$ $Y$s, with $n+m=2$. For $(n,m)=(2,2)$ there are two possible gauge invariant operators, so that we can not determine both using the trace relations. This forces us to introduce an invariant.}
\label{tab:example}
\end{table}

For $n+m=5$, the number of independent single-trace operators for each $(n,m)$ matches the number of independent trace relations. Consequently, no additional invariants are required -- the complete space of gauge-invariant operators is fully determined by the invariants already identified.

\begin{table}[h]
\centering
\begin{tabular}{|c|c|c|c|}
\hline
$(n,m)$ &Independent Trace Relations & Single Trace Operators & New Invariants \\ \hline
$(5,0)$ &$T_3(X^2,X,X,X)=0$ & $\Tr(X^5)$ & none \\ \hline
$(4,1)$ &$T_3(X^2,X,X,Y)=0$ & $\Tr(X^4Y)$ & none \\ \hline
$(3,2)$ &$T_3(Y^2,X,X,X)=0$,& $\Tr(X^3Y^2)$ & none \\
            &$T_3(YX,X,X,Y)=0$ &  $\Tr(X^2YXY)$ & \\ \hline
$(2,3)$ &$T_3(X^2,Y,Y,Y)=0$  & $\Tr(X^2Y^3)$ & none \\
            &$T_3(XY,Y,Y,X)=0$ & $\Tr(XYXY^2)$ & \\ \hline
$(1,4)$ &$T_3(Y^2,Y,Y,X)=0$ & $\Tr(Y^4X)$ & none \\ \hline
$(0,5)$ &$T_3(Y^2,Y,Y,Y)=0$ & $\Tr(Y^5)$ & none \\ \hline
\end{tabular}
\caption{For $n+m=5$ the number of independent single trace operators equals the number of independent trace relations.}
\label{tab:secondexample}
\end{table}

Next we consider the case that $n+m=6$. The table below shows that when $(m,n)=(3,3)$ we again need to introduce one more invariant, and by inspecting (\ref{N3M2}) we know that this is a primary invariant
\bea
S_1=\Tr(XYX^2 Y^2)
\eea
The choice for $S_1$ is not unique.

\begin{table}[h]
\centering
\begin{tabular}{|c|c|c|c|}
\hline
$(n,m)$ &Independent Trace Relations & Single Trace Operators & New Invariants \\ \hline
$(6,0)$ &$T_3(X^3,X,X,X)=0$ & $\Tr(X^6)$ & none \\ \hline
$(5,1)$ &$T_3(X^3,X,X,Y)=0$ & $\Tr(X^5Y)$ & none \\ \hline
$(4,2)$ &$T_3(Y^2,X^2,X,X)=0$ & $\Tr(X^4Y^2)$ & none \\ 
            &$T_2(X,Y,X^3,Y)=0$ & $\Tr(X^3YXY)$ &  \\ 
            &$T_3(XY,XY,X,X)=0$ & $\Tr(X^2YX^2Y)$ & \\ 
\hline
$(3,3)$ &$T_3(X^2,Y^2,X,Y)=0$ & $\Tr(X^3Y^3)$ & $\Tr(XYX^2Y^2)$ \\
            &$T_2(X^2Y,X,Y,Y)=0$ & $\Tr(XYXYXY)$ &  \\
 &$T_3(Y^2X,Y,Y,X)=0$,  & $\Tr(XY^2X^2Y)$ &\\
 & & $\Tr(X^2Y^2XY)$ &\\ \hline
$(2,4)$ &$T_3(Y^2,X^2,Y,Y)=0$ & $\Tr(Y^4X^2)$ & none \\ 
            & $T_3(Y,X,Y^3,X)=0$ & $\Tr(Y^3XYX)$ &  \\ 
 &$T_3(YX,YX,Y,Y)=0$ & $\Tr(Y^2XY^2X)$ &  \\ \hline
$(1,5)$ &$T_3(Y^3,Y,Y,X)=0$ & $\Tr(Y^5X)$ & none \\ \hline
$(0,6)$ &$T_3(Y^3,Y,Y,Y)=0$ & $\Tr(Y^6)$ & none \\ \hline
\end{tabular}
\caption{For $n+m=6$ the number of independent single trace operators equals the number of independent trace relations, except for $n=3=m$ when there is one more single trace operator than the number of indepenent trace relations.}
\label{tab:secondexample}
\end{table}

To give the presentation of the algebra of invariants in terms of $q$-reducible oscillators, multiply the above partition function by $1=\frac{1-x^6}{1-x^6}$ to obtain
\bea
Z(x) &=& \frac{1-x^{12}}{(1-x)^2(1-x^2)^3(1-x^3)^4(1-x^4)(1-x^6)}\label{N3M2}
\eea
We have one extra factor in the denominator implying we have one more oscillator. The numerator specifies that this oscillator obeys a constraint telling us that it is $q$-reducible
\bea
(S_1)^2&=&S_1f_1(\{P_i\})+f_2(\{P_i\})
\eea
The explicit form of this constraint can be found in Appendix C.2 of \cite{deMelloKoch:2025ngs}.

\subsection{$N=4$}\label{Nis4}

The Molien-Weyl partition function for 2 $4\times 4$ matrices is given by \cite{T}
\bea
Z(x)&=&{P_{N=4,M=2}(x)\over (1-x)^2(1-x^2)^3(1-x^3)^4(1-x^4)^6(1-x^6)^2}\label{N4M2}
\eea
where
\bea
P_{N=4,M=2}(x)&=&1 + 2 x^5 + 2 x^6 + 2 x^7 + 4 x^8 + 4 x^9 + 4 x^{10} 
+ 4 x^{11} + 
 2 x^{12} + 4 x^{13} + 4 x^{14}\cr\cr
&& + 4 x^{15} + 4 x^{16} + 2 x^{17}+2 x^{18}+2 x^{19}+x^{24}
\eea
The structure of this Hilbert series indicates that the invariant algebra is generated using 17 primary and 48 secondary invariants. Our objective is to explicitly construct these invariants in order to demonstrate the emergence
$q$-reducible oscillators.

Determining the full set of primary and secondary invariants in this example is a highly nontrivial problem. Conceptually, however, the strategy is straightforward: one must solve the trace relations. These relations immediately imply that all single-trace operators containing fewer than $N+1=5$ matrices are included among the invariants. This yields 15 invariants, all of which are primary. The challenge lies in identifying the remaining 2 primary invariants and the 48 secondary invariants—a formidable technical task. Fortunately, this problem was recently solved in its entirety by mathematicians in \cite{pgeometry}, using methods from noncommutative Poisson geometry. In what follows, we briefly summarize the key ideas of \cite{pgeometry} and then make extensive use of their results.

As stressed in \cite{D2m}, working with traceless matrices significantly simplifies the identities derived from the trace relations. Indeed, for a collection of $N+1$ matrices $W_1,W_2,\cdots,W_{N+1}$ which are each given by arbitrary words in the $X^a$ $a=1,2,\cdots,d$, the trace relations follow by antisymmetrizing the column indices as follows \cite{Pr}
\bea
\sum_{\sigma\in S_N}{\rm sgn}(\sigma)(W_1)_{i_1i_{\sigma(1)}}(W_2)_{i_2 i_{\sigma(2)}}\cdots (W_{N+1})_{i_{N+1}i_{\sigma(N+1)}}=0\label{CayleyHamilton}
\eea
where ${\rm sgn}(\sigma)$ is the parity of $\sigma$. The vanishing of the above identity is clear: antisymmetrizing $N+1$ indices which each take $N$ values always gives zero. If the words are chosen to be traceless, any term in the above sum corresponding to a permutation that contains one cycles, vanishes. The justification for using traceless matrices is due to Procesi~\cite[Section 5]{Pr} who shows an isomorphism of the $\mathbb{Z}^d$-graded algebras
\bea
C_{Nd} \cong \mathbb{C}[u_1,u_2,\dots,u_d] \otimes C_{Nd}(0),
\eea
where $C_{Nd}(0)$ is the algebra of $\mathrm{U}(N)$-invariants of $d$ traceless $N\times N$ matrices. The $u_i$'s are the value of the trace of generic matrices. 

For the algebra $C_{42}$, we choose the following primary invariants:
\bea
\label{primary_invariants}
&& X, Y\qquad A^2, AB, B^2 \qquad A^3, A^2B, AB^2, B^3\qquad
 A^4, A^3B, A^2B^2, AB^3, B^4,\frac12[A,B]^2\cr\cr
&& [A,B]^2A^2, [A,B]^2B^2.
\eea
where we use the traceless matrices
\bea
A= X-\frac14\Tr(X)I_4\qquad B= Y-\frac14\Tr(Y)I_4
\eea
with $I_4$ the $4\times 4$ identity matrix. Notice that the first line of (\ref{primary_invariants}) is simply a list of all loops with no more than 4 matrices in the trace. The second line of (\ref{primary_invariants}) is obtained by using the specific trace relations which determine all single trace operators with 5 or 6 matrices in the trace. We find that all but 2 single trace operators are determined and this forces us to introduce the 2 invariants listed on the last line of (\ref{primary_invariants}).

Next, following \cite{pgeometry}, we equip $C_{N2}$ with a Poisson bracket. This additional structure is useful to uncover relations among trace invariants, as some key relations emerge naturally from Poisson bracket identities. To see how the Poisson bracket is introduced, consider the symplectic form 
\bea
\omega ((X_1,X_2) \, , \, (Y_1,Y_2))=\textrm{Tr}(X_1Y_2-X_2Y_1).
\eea
on pairs of matrices. The symplectic form is ${\mathrm U}(N)$-invariant so this gives the algebra $C_{N2}$ a Poisson algebra structure. To determine the value of the Poisson bracket on specific generators, consider the free noncommutative $\mathbb{C}$-algebra $R$ generated by the two elements $x$ and $y$ and define
\bea
\omega(x,y)=-\omega(y,x)=1, \qquad \omega(x,x)=\omega(y,y)=0.
\eea
There is a Leibniz algebra structure on $R$ defined as follows:
\begin{equation}
\{u_1\cdots u_p, v_1\cdots v_q\}=\sum\limits_{\substack {1\leq i \leq p \\1\leq j \leq q}}\omega(u_i,v_j)u_{i+1}\cdots u_pu_1\cdots u_{i-1}v_{j+1}\cdots v_q v_1\cdots v_{j-1},
\end{equation}
where $u_1,\dots,u_p,v_1,\dots, v_q$ are either $x$ or $y$. Define the subspace $\textrm{Com}(R)=\textrm{Span}\{ab-ba \mid a,b\in R\}$ which forms a central ideal. The quotient algebra $\mathcal{N}=R/\text{Com}(R)$, inherits the structure of a Lie algebra as shown in~\cite{G}. The trace map
\begin{equation}\nonumber
\begin{split}
tr\colon {\cal N}&\rightarrow C_{N2}\\
x^{k_1}y^{l_1}\cdots x^{k_m}y^{l_m}&\mapsto \big( (X,Y)\mapsto \mathrm{Tr}(X^{k_1}Y^{l_1}\cdots X^{k_m}Y^{l_m})\big)
\end{split}
\end{equation}
is a well-defined Lie algebra homomorphism. Using it we can compute the brackets on $C_{N2}$, turning it into a Poisson algebra.

The equivalence class of cyclically equivalent words is called a necklace. The degree lexicographic order on two-letter words can be used to induce an order on $\mathcal{N}$: necklaces are compared by choosing a canonical representative from each equivalence class -- the minimal element with respect to degree lexicographic order -- and ordering accordingly. Within the full set of necklaces, particular importance is given to breaking pairs, defined below. Poisson brackets between breaking pairs yield non trivial relations, needed to identify and eliminate redundancies among invariants. These relations \emph{break} the redundancy among invariants by producing expressions where a more complex secondary invariant is related to simpler ones, ultimately reducing the space to a minimal, independent set.

We say that a necklace is $CH_N$ if at least one of the words in its equivalence class contains the same subword $N$ times consecutively. Necklaces which are $CH_N$ are easily determined by trace relations where $N$ of the words chosen in (\ref{CayleyHamilton}) are taken to be equal to the repeated subword\footnote{The identity (\ref{CayleyHamilton}) is a generalization of the Cayley-Hamilton identity. The ``$CH$'' in the name $CH_N$ is for Cayley-Hamilton.}. 

For a fixed bidegree $(r,s)$, a pair of necklaces $(w_1,w_2)$ is called a breaking pair if it satisfies all of the following:
\begin{itemize}
\item[1.] Degree sum: Their bidegrees add up to $(r+1,s+1)$.
\item[2.] Order condition: The first necklace $w_1$ must have (total) degree at least 2, and must be smaller than $w_2$ in degree‑lexicographic order.
\item[3.] Special case when first degree = 2: In that case, $w_2$ must be a $CH_N$‑necklace.
\item[4.] Generator‑trace condition: At least one of the two necklaces must have a trace that is not among the minimal generators (i.e., its trace does not appear as a linear term in the minimal generating set).
\end{itemize}
These pairs are used to compute the value of algebraic expressions of the traces of necklaces that are not $CH_N$. For example, for $N=4$ in bidegree~$(3, 2)$ the pair $(B^2, A^4B)$ is a breaking pair, but the pair $(A^2B, ABAB)$ is not, since~$\Tr(A^2B)$ is exactly $P_7$ and $\Tr(ABAB)$ appears in the linear expansion of~$P_{15}$.

In summary, breaking pairs are special pairs of necklaces chosen to systematically generate equations which help express the traces of complex necklaces in terms of known generators, especially in cases where direct application of the trace relations is tedious.

In Appendix \ref{AlgExec} we give the details of how this algorithm is executed. It produces the following primary invariants
\bea
P_1&=&\Tr(X)\qquad P_2\,\,=\,\,\Tr(Y)\qquad P_3\,\,=\,\,\Tr(A^2)\qquad P_4\,\,=\,\,\Tr(AB)\cr\cr
P_5&=&\Tr(B^2)\qquad P_6\,\,=\,\,\Tr(A^3)\qquad P_7\,\,=\,\,\Tr(A^2B)\qquad P_8=\Tr(AB^2)\cr\cr
P_9&=&\Tr(B^3)\qquad P_{10}\,\,=\,\,\Tr(A^4)\qquad P_{11}\,\,=\,\,\Tr(A^3B)\qquad P_{12}\,\,=\,\,\Tr(A^2B^2)\cr\cr  P_{13}&=&\Tr(AB^3)\qquad P_{14}\,\,=\,\,\Tr(B^4)\qquad P_{15}=\frac{1}{2}\Tr([A,B]^2)\qquad P_{16}=\Tr([A,B]^2A^2)\cr\cr 
P_{17}&=&\Tr([A,B]^2B^2)
\eea
the following secondary invariants
\bea
S_0&=&1\qquad S_1\,\,=\,\,\Tr([A,B]^2A)\qquad S_2=\Tr([A,B]^2B)\qquad S_3=\Tr([A,B]^2(AB+BA))\cr\cr
S_4&=&\frac13\Tr([A,B]^3)\qquad S_5\,\,=\,\,\Tr([A,B]^3A)\qquad S_6\,\,=\,\,\Tr([A,B]^3B)\cr\cr
S_7&=&\Tr([A,B]^3A^2)\qquad S_8\,\,=\,\,\frac12\Tr([A,B]^3(AB+BA))\qquad
S_9\,\,=\,\,\Tr([A,B]^3B^2)\cr\cr 
S_{10}&=&\frac12\Tr([A,B]^4)
S_{11}\,\,=\,\,\Tr([A,B]^3A^3)\qquad S_{12}=\frac13\Tr([A,B]^3(A^2B+ABA+BA^2))\cr\cr
S_{13}&=&\frac13\Tr([A,B]^3(AB^2+BAB+B^2A))\qquad S_{14}=\Tr([A,B]^3B^3)\cr\cr
S_{15}&=&\Tr([A,B]^3(A^2B^2-AB^2A-BA^2B+B^2A^2))
\eea
and the following multiparticle secondary invariants
\bea
S_{16}&=&S_1^2\qquad S_{17}\,\,=\,\,S_1 S_2\qquad S_{18}\,\,=\,\,S_2^2\qquad
S_{19}\,\,=\,\,S_1 S_3\qquad S_{20}\,\,=\,\,S_1S_4\cr\cr
S_{21}&=&S_2 S_3\qquad S_{22}\,\,=\,\,S_2 S_4\qquad S_{23}\,\,=\,\,S_3 S_4\qquad S_{24}\,\,=\,\,S_4^2\qquad
S_{25}\,\,=\,\,S_3 S_5\cr\cr
S_{26}&=&S_3 S_6\qquad S_{27}\,\,=\,\,S_4 S_5\qquad S_{28}\,\,=\,\,S_4 S_6\qquad
S_{29}\,\,=\,\,S_3 S_7\qquad S_{30}\,\,=\,\,S_3 S_9\cr\cr
S_{31}&=&S_4 S_{10}\qquad S_{32}\,\,=\,\,S_5 S_6\qquad S_{33}\,\,=\,\,S_5 S_7\qquad S_{34}\,\,=\,\,S_6 S_7
\qquad S_{35}\,\,=\,\,S_6 S_8\cr\cr
S_{36}&=&S_6 S_9\qquad S_{37}\,\,=\,\,S_7 S_8\qquad S_{38}\,\,=\,\,S_8 S_9\qquad S_{39}\,\,=\,\,S_8 S_{10}\qquad S_{40}\,\,=\,\,S_{10}^2\cr\cr
S_{41}&=&S_9 S_{11}\qquad S_{42}\,\,=\,\,S_9 S_{12}\qquad S_{43}\,\,=\,\,S_{10} S_{15}\qquad S_{44}\,\,=\,\,S_{12} S_{13}\qquad S_{45}\,\,=\,\,S_4^2 S_5\cr\cr 
S_{46}&=&S_4^2 S_6\qquad S_{47}\,\,=\,\,S_8 S_{10}^2
\eea

Looking at the multi-particle secondary invariants, it is clear that a number of $q$-reducible oscillators have appeared. For example, both $S_1$ and $S_2$, as well as $S_1^2$ and $S_2^2$ appear, but $S_1^3$ and $S_2^3$ do not.
In summary, the collective field theory for N=4 consists of:
17 primary oscillators $P_i$ and 15 secondary oscillators $S_i$.On the
Fock space the primary oscillators act freely,while the secondary ones act in a q-reduced manner summarized in Eq.4.19.This implies a 
growing reduction at higher energy  levels.

Although the description of the invariants becomes much more complicated as $N$ increases, simply because of the explosion in the number of secondary invariants, a number of general features are clear. First, for an operator to become redundant by trace relations, it must be constructed using more than $N$ matrices. Thus, if a single trace secondary invariant has $k$ matrices in the trace, it can be raised to a power $q$ with $q\ge \frac{N}{k}$. 
For $S_1$ and $S_2$ above, $k=5$, $N=4$ and $q=3$, so that the inequality is obeyed. As $N$ is increased the number of $q$-reducible oscillators increases, as does the possible range of values for $q$.

Another feature that our small $N$ example does not illustrate clearly, is that for moderately large values of $N$, the number of $q$-reducible oscillators is a tiny fraction of the total number of secondary invariants. This is most clearly seen by noting that the $q$-reducible oscillators have a length of most of order $N$, so the order of the set they are selected from grows exponentially in $N$. This is much slower than the growth in the number of secondary invariants which grows as the exponential of $N^2$.
\section{Discussion}\label{discussion}

We have analyzed the structure of the space of secondary invariants in several explicit examples. This space contains operators constructed from $O(N^2)$ fields, which we interpret as corresponding to new spacetime geometries in the holographic dual, as well as operators built from $O(N)$ fields, which we associate with solitonic objects such as D-branes. In addition, there exists a class of operators constructed from $k\ll N$ fields, which has been the main focus of our study in this paper. We have presented evidence that these correspond to $q$-reducible oscillators.

Through explicit construction, we find the secondary invariants that are single-trace operators built from $k\ll N$ fields. We refer to these as primitive secondary invariants. Powers of these primitive operators also appear as distinct secondary invariants. However, there exists a maximum power to which a given primitive secondary invariant can be raised while still contributing a new independent invariant. Beyond this maximal power, higher powers no longer yield new operators -- they are eliminated from the spectrum due to the trace relations, which allow them to be expressed in terms of already existing operators.

This saturation phenomenon implies that, at low energies, these primitive secondary invariants behave as conventional perturbative degrees of freedom. But when excited beyond the maximal power, they cease to be independent states. As we have discussed, this behavior is a direct consequence of the trace relations, which encode the finite-$N$ structure of the operator algebra.

Our results reveal a striking mechanism in quantum gravity whereby an apparent proliferation of degrees of freedom in the ultraviolet is sharply curtailed by trace relations in the holographic dual. Our explicit construction provides a concrete illustration of how gravity enforces such truncations: in the dual gauge theory, trace identities eliminate redundant states. This mechanism highlights a deep and perhaps universal feature of quantum gravity -- that the true number of independent observables is far smaller than naïvely expected from semiclassical considerations.

\begin{center} 
{\bf Acknowledgements}
\end{center}
We would like to thank Jo\~ao Rodrigues for discussions. 
The work of RdMK is supported by a start up research fund of Huzhou University, a Zhejiang Province talent award and by a Changjiang Scholar award. The work of A.J. is supported by the U.S. Department of Energy under contract DE-SC0010010.

\begin{appendix}
\section{Invariants for the $N=3$, $d=2$ boson model}\label{DeriveGens}

In this Appendix we will work out the generators for 3 bosons in $d=2$ dimensions. To simplify the analysis, we introduce the notation
\bea
(n,m)&\equiv& \sum_{i=1}^3(x_i)^n (y_i)^m
\eea
Our claim is that the primary and secondary invariants are given by
\bea
P_1&=&(1,0)\qquad\qquad P_2\,\,=\,\,(2,0)\qquad\qquad P_3\,\,=\,\,(3,0)\cr\cr
P_4&=&(0,1)\qquad\qquad P_5\,\,=\,\,(0,2)\qquad\qquad P_6\,\,=\,\,(0,3)\cr\cr
S_1&=&(1,1)\qquad\qquad S_2\,\,=\,\,(2,1)\qquad\qquad S_3\,\,=\,\,(1,2)\cr\cr
S_4&=&(1,1)^2\qquad\qquad S_5\,\,=\,\,(1,1)^3\qquad\qquad S_0=1
\eea
To prove the claim, we must verify that the complete set of single trace invariants can be written in terms of the above generators. To demonstrate this we need to use the trace relations $T_2(A,B,C)=0$ where $T_2(A,B,C)$ is defined in (\ref{tracerelationN2}). We will also make use of the matrices $X$ and $Y$ with $X,Y$ defined as in (\ref{matsdefined}). First, every invariant of degree 3 or less must be included amongst the generating set because the trace relations only start at degree 4.

\subsection{Degree $4$}

There are 5 single trace invariants, all of which are determined by the trace relations, as summarized below:
\begin{itemize}
\item $(4,0):$ Use the trace relation
\bea
-6\Tr(X^4)+8\Tr(X)\Tr(X^3)+3\Tr(X^2)^2-6\Tr(X^2)\Tr(X)^2+\Tr(X^4)&=&0\cr
&&
\eea
to find
\bea
-6(4,0)+8(1,0)(3,0)+3(2,0)^2-6(2,0)(1,0)^2+(1,0)^4&=&0
\eea
Thus
\bea
(4,0)&=&\frac{1}{6}\Big((P_1)^4-6P_2 (P_1)^2+3(P_2)^2+8P_1 P_3\Big)
\eea
\item $(3,1):$ Use the trace relation
\bea
&&-6 \Tr(YX^3)+2\Tr(Y)\Tr(X^3)+6\Tr(X)\Tr(YX^2)+3\Tr(XY)\Tr(Y^2)\cr\cr
&&-3\Tr(X)^2\Tr(XY)-3\Tr(X)\Tr(Y)\Tr(X^2)+\Tr(Y)\Tr(X)^3\,\,=\,\,0
\eea
to find
\bea
&&-6(3,1)+2(0,1)(3,0)+6(1,0)(2,1)+3(1,1)(2,0)-3(1,0)^2(1,1)\cr\cr
&&-3(0,1)(1,0)(2,0)+(0,1)(1,0)^3\,\,=\,\,0
\eea
Thus
\bea
(3,1)&=&\frac{1}{6}\Big(2P_4 P_3+6P_1 S_2+3S_1 P_2-3(P_1)^2 S_1-3P_4 P_1 P_2+P_4 (P_1)^3\Big)
\eea
This is linear in the secondary invariants.
\item $(2,2):$ Use the trace relation
\bea
&&-2\Tr(XYXY)-4\Tr(X^2Y^2)+4\Tr(Y)\Tr(YX^2)-4\Tr(X)\Tr(XY^2)\cr\cr
&&+\Tr(X^2)\Tr(Y^2)+2\Tr(XY)^2-\Tr(X)^2\Tr(Y^2)-\Tr(Y)^2\Tr(X^2)\cr\cr
&&-4\Tr(X)\Tr(Y)\Tr(XY)+\Tr(X)^2\Tr(Y)^2\,\,=\,\,0
\eea
to find
\bea
&&-6(2,2)+4(0,1)(2,1)+4(1,0)(1,2)+(2,0)(0,2)+2(1,1)^2\cr\cr
&&-(1,0)^2(0,2)-(0,1)^2 (2,0)-4(1,0)(0,1)(1,1)+(1,0)^2(0,1)^2\,\,=\,\,0\cr
&&
\eea
Note that in order that we are able to solve for $(2,2)$, we are forced to include the secondary invariant $S_4=(1,1)^2$. Thus
\bea
(2,2)&=&\frac{1}{6}\Big(4P_4 S_2+4P_1 S_3+P_2 P_5+2S_4-(P_1)^2P_5-(P_4)^2P_2\cr\cr
&&-4P_1 P_4 S_1+(P_1)^2(P_4)^2\Big)
\eea

\item $(1,3):$ Argue as for $(3,1)$ to obtain
\bea
(1,3)&=&\frac{1}{6}\Big(2P_1 P_6+6P_4 S_3+3S_1 P_5-3P_4^2 S_1-3P_1 P_4 P_5+P_1P_4^3\Big)
\eea
\item $(0,4):$ Argue as for $(4,0)$ to obtain
\bea
(0,4)&=&\frac{1}{6}\Big((P_4)^4-6P_5 (P_4)^2+3(P_5)^2+8P_4 P_6\Big)
\eea
\end{itemize}

Thus, we have determined all invariants of degree 4 and we have determined the form of the secondary invariant $S_4$.

\subsection{Degree $5>N$}

There are 6  single trace invariants given by
\begin{itemize}
\item $(5,0):$ From the trace relation
\bea
&&-6\Tr(X^5)+5\Tr(X^2)\Tr(X^3)+6\Tr(X)\Tr(X^4)-3\Tr(X)^2\Tr(X^3)-3\Tr(X^2)^2\Tr(X)\cr\cr
&&+\Tr(X^2)\Tr(X)^3\,\,=\,\,0
\eea
we find
\bea
(5,0)&=&\frac{1}{6}\Big(5(2,0)(3,0)+6(1,0)(4,0)-3(1,0)^2(3,0)-3(2,0)^2(1,0)+(2,0)(3,0)\Big)\cr
&&
\eea
Thus
\bea
(5,0)&=&\frac{1}{6}\Big(5P_2 P_3+(P_1)^5-5P_2 (P_1)^3+5(P_1)^2 P_3\Big)
\eea
\item $(4,1):$ From the trace relation
\bea
&&-6\Tr(YX^4)+2\Tr(XY)\Tr(X^3)+6\Tr(X)\Tr(X^3Y)+3\Tr(X^2Y)\Tr(X^2)\cr\cr
&&-3\Tr(X)^2\Tr(X^2Y)-3\Tr(XY)\Tr(X)\Tr(X^2)+\Tr(XY)\Tr(X^3)\,\,=\,\,0\cr
&&
\eea
we find
\bea
(4,1)&=&\frac{1}{6}\Big(2(1,1)(3,0)+6(1,0)(3,1)+3(2,1)(2,0)-3(1,0)^2(2,1)\cr\cr
&&-3(1,1)(1,0)(2,0)+(1,1)(1,0)^3\Big)
\eea
Thus
\bea
(4,1)&=&\frac{1}{6}\Big(2S_1 P_3+2P_4 P_3 P_1 +3(P_1)^2 S_2-2(P_1)^3 S_1-3P_4 (P_1)^2 P_2+P_4 (P_1)^4\cr\cr
&&+3S_2 P_2\Big)
\eea
\item $(3,2):$ From the trace relation
\bea
&&-6\Tr(Y^2X^3)+2\Tr(Y^2)\Tr(X^3)+6\Tr(X)\Tr(X^2Y^2)+3\Tr(XY^2)\Tr(X^2)\cr\cr
&&-3\Tr(X)^2\Tr(Y^2X)-3\Tr(Y^2)\Tr(X)\Tr(X^2)+\Tr(Y^2)\Tr(X)^3\,\,=\,\,0
\eea
we find
\bea
(3,2)&=&\frac{1}{6}\Big(2(0,2)(3,0)+6(1,0)(2,2)+3(1,2)(2,0)-3(1,0)^2(1,2)\cr\cr
&&\qquad-3(0,2)(1,0)(2,0)+(0,2)(1,0)^3\Big)
\eea
Thus
\bea
(3,2)&=&\frac{1}{6}\Big(2P_5 P_3+4P_4 P_1 S_2 +4(P_1)^2 S_3+2P_1 S_4-(P_4)^2 P_2 P_1-4(P_1)^2P_4 S_1\cr\cr
&&\qquad+(P_1)^3(P_4)^2+3S_3 P_2 -3(P_1)^2 S_3-2P_5 P_1 P_2 \Big)
\eea
\item $(2,3):$ Arguing as we did for $(3,2)$ we find
\bea
(2,3)&=&\frac{1}{6}\Big(2P_2 P_6+4P_1 P_4 S_3 +4(P_4)^2 S_2 +2P_4 S_4-(P_1)^2 P_5 P_4-4(P_4)^2P_1 S_1\cr\cr
&&\qquad+(P_4)^3(P_1)^2+3S_2 P_5 -3(P_4)^2 S_2-2P_2 P_4 P_5 \Big)
\eea
\item $(1,4):$ Arguing as we did for $(4,1)$ we find
\bea
(1,4)&=&\frac{1}{6}\Big(2S_1 P_6+2P_1 P_6 P_4 +3(P_4)^2 S_3-2(P_4)^3 S_1-3P_1 (P_4)^2 P_5+P_1 (P_4)^4\cr\cr
&&+3S_3 P_5\Big)
\eea
\item $(0,5):$ Arguing as for $(5,0)$ we have
\bea
(0,5)&=&\frac{1}{6}\Big(5P_5 P_6+(P_4)^5-5P_5 (P_4)^3+5(P_4)^2 P_6\Big)
\eea
\end{itemize}
In every expression, the solitons appear at most linearly.

\subsection{Degree $6>N$}

There are 7 single trace invariants given by
\begin{itemize}
\item $(6,0):$ From the trace relation
\bea
 &&-6\Tr(X^6)+2\Tr(X^3)^2+6\Tr(X)\Tr(X^5)+3\Tr(X^4)\Tr(X^2)-3\Tr(X)^2\Tr(X^4)\cr\cr
 &&-3\Tr(X^3)\Tr(X^2)\Tr(X)+\Tr(X^3)\Tr(X)^3=0
 \eea
 we find
 \bea
 (6,0)&=&\frac{1}{12} \Big((P_1)^6-3 (P_1)^4 P_2-9 (P_1)^2 (P_2)^2+3 (P_2)^3+4 (P_1)^3 P_3\cr\cr
 &&+12 P_1 P_2 P_3+4 (P_3)^2\Big)
 \eea
\item $(5,1):$ From the trace relation
\bea
&&-6\Tr(YX^5)+2\Tr(YX^2)\Tr(X^3)+6\Tr(X)\Tr(X^4Y)+3\Tr(YX^3)\Tr(X^2)\cr\cr
&&-3\Tr(X)^2\Tr(YX^4)-3\Tr(YX^2)\Tr(X^2)\Tr(X)+\Tr(YX^2)\Tr(X)^3=0
\eea
we find
\bea
(5,1)&=&\frac{1}{12}\Big(-(P_1)^4 S_1-6 (P_1)^2 S_1 P_2+3 S_1 (P_2)^2+2 (P_1)^3 S_2+4 S_2 P_3\cr\cr
&&+P_4 ((P_1)^2+P_2) ((P_1)^3-3 P_1 P_2+2 P_3)+P_1 (6 P_2 S_2+4 S_1 P_3)\Big)\cr
&&
\eea
\item $(4,2):$ From the trace relation
\bea
&&-6\Tr(Y^2X^4)+2\Tr(Y^2X)\Tr(X^3)+6\Tr(X)\Tr(X^3Y^2)+3\Tr(X^2Y^2)\Tr(X^2)\cr\cr
&&-3\Tr(X)^2\Tr(Y^2X^2)-3\Tr(Y^2X)\Tr(X)\Tr(X^2)+\Tr(Y^2X)\Tr(X)^3=0
\eea
we find
\bea
(4,2)&=&\frac{1}{12}\Big((P_4)^2 ((P_1)^4 - (P_2)^2) - 4 P_4 ((P_1)^2+P_2) (P_1 S_1-S_2)\cr\cr
&&+P_5((P_1)^4-6(P_1)^2 P_2+(P_2)^2 + 4P_1 P_3)+2(P_1)^2 S_4 + 2S_4 P_2\cr\cr
&&+ 4 P_1 S_3 P_2+4 S_3 P_3\Big)
\eea
\item $(3,3):$ Using the trace relation 
\bea
&&-6\Tr(X^3Y^3)+2\Tr(X^3)\Tr(Y^3)+6\Tr(X)\Tr(X^2Y^3)+3\Tr(XY^3)\Tr(X^2)\cr\cr
&&-3\Tr(X)^2\Tr(Y^3X)-3\Tr(Y^3)\Tr(X)\Tr(X^2)+\Tr(Y^3)\Tr(X)^3=0
\eea
we find
\bea
(3,3)&=&\frac{1}{12}\Big((P_4)^3 (P_1)^3+P_4 P_5 (P_1)^3-5 (P_4)^2 (P_1)^2 S_1-3 P_5(P_1)^2 S_1+4P_4 P_1 S_4\cr\cr
&&+2 P_4 (P_1)^2 S_3+(P_4)^3 P_1 P_2-7 P_4 P_5 P_1 P_2-3 (P_4)^2 S_1 P_2\cr\cr
&&+3 P_5 S_1 P_2+6 P_4 S_3 P_2+2 (P_4)^2 P_1 S_2+6 P_5 P_1 S_2+4 P_6 P_3\Big)
\eea
\item $(2,4):$ Arguing as we did for $(4,2)$ we find
\bea
(2,4)&=&\frac{1}{12}\Big((P_1)^2 ((P_4)^4 - (P_5)^2) - 4 P_1 ((P_4)^2+P_5) (P_4 S_1-S_3)\cr\cr
&&+P_2((P_4)^4-6(P_4)^2 P_5+(P_5)^2 + 4P_4 P_6)+2(P_4)^2 S_4 + 2S_4 P_5\cr\cr
&&+ 4 P_4 S_2 P_5+4 S_2 P_6\Big)
\eea
\item $(1,5):$ Arguing as we did for $(5,1)$ we find
\bea
(1,5)&=&\frac{1}{12}\Big(-(P_4)^4 S_1-6 (P_4)^2 S_1 P_5+3 S_1 (P_5)^2+2 (P_4)^3 S_3+4 S_3 P_6\cr\cr
&&+P_1 ((P_4)^2+P_5) ((P_4)^3-3 P_4 P_5+2 P_6)+P_4 (6 P_5 S_3+4 S_1 P_6)\Big)
\eea
\item $(0,6):$ Arguing exactly as for $(6,0)$ we find
 \bea
 (0,6)&=&\frac{1}{12} \Big((P_4)^6-3 (P_4)^4 P_5-9 (P_4)^2 (P_5)^2+3 (P_5)^3+4 (P_4)^3 P_6\cr\cr
 &&+12 P_4 P_5 P_6+4 (P_6)^2\Big)
 \eea
\end{itemize}

Notice that these relations still have not forced us to introduce the final secondary invariant $S_5$. To see why it is needed, consider the following trace relation
\bea
&&-2\Tr(XYXYXY)-2\Tr(XYX^2Y^2)-2\Tr(X^2YXY^2)+2\Tr(X)\Tr(XYXY^2)\cr\cr
&&+2\Tr(Y)\Tr(X^2YXY)+3\Tr(XY)\Tr(XYXY)+2\Tr(XY)\Tr(X^2Y^2)+2\Tr(X^2Y)\Tr(XY^2)\cr\cr
&&-\Tr(XYXY)\Tr(X)\Tr(Y)-2\Tr(X^2Y)\Tr(XY)\Tr(Y)-2\Tr(XY^2)\Tr(XY)\Tr(X)\cr\cr
&&-\Tr(XY)^3+\Tr(X)\Tr(Y)\Tr(XY)^2=0
\eea
which can be rewritten to obtain (note that now $S_5$ participates - so we have introduced the last secondary invariant)
\bea
&&-2 P_1 P_2 P_4^3 - 2 P_1^3 P_4 P_5 + 6 P_1 P_2 P_4 P_5 + 2 P_3 P_4 P_5 +2 P_1 P_2 P_6 - 6 P_3 P_6 + 4 P_1^2 P_4^2 S_1\cr\cr
 &&+ 2 P_2 P_4^2 S_1+ 2 P_1^2 P_5 S_1 - 2 P_2 P_5 S_1 - 7 P_1 P_4 S_1^2 + 4 P_4 S_1 S_2 + 4 P_1 S_1 S_3 - 
 6 P_1 P_5 S_2 + 6 S_2 S_3\cr\cr
 &&- 6 P_2 P_4 S_3 - 3 P_1 P_4 S_4 + 2 S_5=0
\label{frst}
\eea
as well as the trace relation
\bea
&&-2\Tr(X^3Y^3)-2\Tr(X^2YXY^2)-2\Tr(X^2Y^2XY)+2\Tr(X^2)\Tr(XY^3)+2\Tr(XY)\Tr(X^2Y^2)\cr\cr
&&+2\Tr(Y)\Tr(X^3Y^2)+2\Tr(Y)\Tr(X^3Y^2)+2\Tr(Y)\Tr(XYX^2Y)+\Tr(X^3Y)\Tr(Y^2)\cr\cr
&&+2\Tr(X^2Y)\Tr(Y^2X)-\Tr(X^2)\Tr(XY)\Tr(Y^2)-2\Tr(X^2)\Tr(Y)\Tr(XY^2)\cr\cr
&&-2\Tr(XY)\Tr(Y)\Tr(YX^2)-\Tr(Y)^2\Tr(X^3Y)+\Tr(X^2)\Tr(XY)\Tr(Y)^2=0
\eea
which can be rewritten to obtain (note that again $S_5$ participates)
\bea
&&-P_4 P_5 (P_1)^3+2 (P_4)^2 (P_1)^2 S_1+2 P_5 (P_1)^2 S_1-6 P_4 P_1 S_4+2 S_5-P_4 (P_1)^2 S_3+4 P_1 S_1 S_3\cr\cr
&&-(P_4)^3 P_1 P_2+2 P_4 P_5 P_1 P_2+2 P_6 P_1 P_2+2 (P_4)^2 S_1 P_2-2 P_5 S_1 P_2-3 P_4 S_3 P_2\cr\cr
&&+2 (P_4)^2 P_1 S_2-6 P_5 P_1 S_2-2 P_4 S_1 S_2+6 S_2 S_3-(P_4)^3 P_3+5 P_4 P_5 P_3-6 P_6 P_3=0\cr
&&\label{scnd}
\eea
By swapping $X$ and $Y$ in the above argument we obtain the relation (note that again $S_5$ participates)
\bea
&&-P_1 P_2 (P_4)^3+2 (P_1)^2 (P_4)^2 S_1+2 P_2 (P_4)^2 S_1-6 P_1 P_4 S_4+2 S_5-P_1 (P_4)^2 S_2+4 P_4 S_1 S_2\cr\cr
&&-(P_1)^3 P_4 P_5+2 P_1 P_2 P_4 P_5+2 P_3 P_4 P_5+2 (P_1)^2 S_1 P_5-2 P_2 S_1 P_5-3 P_1 S_2 P_5\cr\cr
&&+2 (P_1)^2 P_4 S_3-6 P_2 P_4 S_3-2 P_1 S_1 S_3+6 S_3 S_2-(P_1)^3 P_6+5 P_1 P_2 P_6-6 P_3 P_6=0\cr
&&\label{thrd}
\eea
Note that now (\ref{frst}), (\ref{scnd}) and (\ref{thrd}) can be solved to obtain expressions for $S_1 S_2$, $S_1 S_3$ and $S_2 S_3$. To do this however, we had to introduce $S_5$. The result is
\bea
S_1S_2&=&\frac{1}{6} \Big(-P_3 P_4^2+P_1^3 P_5+3 P_3 P_5+3 P_2 S_3-P_1^2 (2 P_4 S_1+S_3)+P_1 (P_2 (P_4^2-4 P_5)\cr\cr
&&+7 S_1^2+2 P_4 S_2-3 S_4)\Big)\,\,\equiv\,\, f_{12}(\{S_a,P_i\})\cr\cr
S_1S_3&=&\frac{1}{6} \Big(P_1^2 (P_4 P_5-P_6)+P_2 (P_4^3-4 P_4 P_5+3 P_6)+7 P_4 S_1^2-P_4^2 S_2+3 P_5 S_2\cr\cr
&&+2 P_1 P_4 (-P_4 S_1+S_3)-3 P_4 S_4\Big)\,\,\equiv\,\, f_{13}(\{S_a,P_i\})\cr\cr
S_2S_3&=&\frac{1}{18}\Big(2 P_1^3 (P_4 P_5+P_6)+2 P_3 (P_4^3-6 P_4 P_5+9 P_6)-2 P_1^2 (2 P_4^2 S_1+3 P_5 S_1+P_4 S_3)\cr\cr
&&+P_1 (2 P_2 (P_4^3-P_4 P_5-6 P_6)-7 P_4 S_1^2-2 P_4^2 S_2+12 P_5 S_2+21 P_4 S_4)\cr\cr
&&-6 (P_2 P_4^2 S_1-P_2 P_5 S_1-2 P_2 P_4 S_3+S_5)\Big)\,\,\equiv\,\, f_{23}(\{S_a,P_i\})
\eea
It is satisfying - and non-trivial - that we obtain a polynomial expression for each of the quadratic combinations above after solving (\ref{frst}), (\ref{scnd}) and (\ref{thrd}). Thus, this analysis has determined the primary and secondary invariants for a system of $N=3$ bosons in $d=2$ dimensions.

Finally, we record the following constraints
\bea
S_1 S_1&=&S_4\,\,\equiv\,\,f_{11}(\{S_a,P_i\})
\eea
\bea
S_1 S_4&=&S_5\,\,\equiv\,\,f_{14}(\{S_a,P_i\})
\eea
\bea
S_2^2&=&\frac{1}{6} \Big(P_5 P_1^4+P_4^2 P_1^2 P_2-4 P_5 P_1^2 P_2-P_4^2 P_2^2+P_5 P_2^2+2 P_5 P_1 P_3-2 P_4 P_1^3 S_1+2 P_4 P_1 P_2 S_1\cr\cr
&&-4 P_4 P_3 S_1+3 P_1^2 S_4-3 P_2 S_4-2 P_1 P_2 S_3+6 P_3 S_3+4 P_4 P_2 S_2+2 P_2 S_4\Big)\cr\cr
&\equiv& f_{22}(\{S_a,P_i\})
\eea
\bea
S_3^2&=&\frac{1}{6} \Big(P_2 P_4^4+P_1^2 P_4^2 P_5-4 P_2 P_4^2 P_5-P_1^2 P_5^2+P_2 P_5^2+2 P_2 P_4 P_6-2 P_1 P_4^3 S_1+2 P_1 P_4 P_5 S_1\cr\cr
&&-4 P_1 P_6 S_1+3 P_4^2 S_4-3 P_5 S_4-2 P_4 P_5 S_2+6 P_6 S_2+4 P_1 P_5 S_3+2 P_5 S_4\Big)\cr\cr
&\equiv& f_{33}(\{S_a,P_i\})
\eea
\bea
S_2 S_4&=&\frac{1}{36}\Big(P_1^2 P_2 P_4^3+3 P_2^2 P_4^3-2 P_1 P_3 P_4^3+P_1^4 P_4 P_5-P_1^2 P_2 P_4 P_5-12 P_2^2 P_4 P_5\cr\cr
&&+6 P_1 P_3 P_4 P_5+P_1^4 P_6-6 P_1^2 P_2 P_6+9 P_2^2 P_6-2 P_1^3 P_4^2 S_1-6 P_3 P_4^2 S_1\cr\cr
&&+6 P_1^3 P_5 S_1-24 P_1 P_2 P_5 S_1+18 P_3 P_5 S_1-14 P_1^2 P_4 S_1^2+21 P_2 P_4 S_1^2+5 P_1^2 P_4^2 S_2\cr\cr
&&-3 P_2 P_4^2 S_2-3 P_1^2 P_5 S_2+9 P_2 P_5 S_2-4 P_1^3 P_4 S_3+12 P_1 P_2 P_4 S_3+6 P_1^2 P_4 S_4\cr\cr
&&-9 P_2 P_4 S_4+24 P_1 S_5\Big)\cr\cr
&\equiv& f_{24}(\{S_a,P_i\})
\eea
\bea
S_3 S_4&=&\frac{1}{36}\Big(P_4^2 P_5 P_1^3+3 P_5^2 P_1^3-2 P_4 P_6 P_1^3+P_4^4 P_1 P_2-P_4^2 P_5 P_1 P_2-12 P_5^2 P_1 P_2\cr\cr
&&+6 P_4 P_6 P_1 P_2+P_4^4 P_3-6 P_4^2 P_5 P_3+9 P_5^2 P_3-2 P_4^3 P_1^2 S_1-6 P_6 P_1^2 S_1+6 P_4^3 P_2 S_1\cr\cr
&&-24 P_4 P_5 P_2 S_1+18 P_6 P_2 S_1-14 P_4^2 P_1 S_4+21 P_5 P_1 S_4+5 P_4^2 P_1^2 S_3-3 P_5 P_1^2 S_3\cr\cr
&&-3 P_4^2 P_2 S_3+9 P_5 P_2 S_3-4 P_4^3 P_1 S_2+12 P_4 P_5 P_1 S_2+6 P_4^2 P_1 S_4-9 P_5 P_1 S_4+24 P_4 S_5\Big)\cr\cr
&\equiv& f_{34}(\{P_i\})
\eea
\bea
S_1^4&=&\frac{1}{36} \Big(-P_1^2 P_2 P_4^4+3 P_2^2 P_4^4-4 P_1 P_3 P_4^4-P_1^4 P_4^2 P_5+P_1^2 P_2 P_4^2 P_5-12 P_2^2 P_4^2 P_5\cr\cr
&&+18 P_1 P_3 P_4^2 P_5+3 P_1^4 P_5^2-12 P_1^2 P_2 P_5^2-9 P_2^2 P_5^2+18 P_1 P_3 P_5^2-4 P_1^4 P_4 P_6\cr\cr
&&+18 P_1^2 P_2 P_4 P_6+18 P_2^2 P_4 P_6-36 P_1 P_3 P_4 P_6+2 P_1^3 P_4^3 S_1+6 P_1 P_2 P_4^3 S_1+12 P_3 P_4^3 S_1\cr\cr
&&+6 P_1^3 P_4 P_5 S_1+6 P_1 P_2 P_4 P_5 S_1-72 P_3 P_4 P_5 S_1+12 P_1^3 P_6 S_1-72 P_1 P_2 P_6 S_1+108 P_3 P_6 S_1\cr\cr
&&+28 P_1^2 P_4^2 S_4+84 P_2 P_4^2 S_4-126 P_1^2 P_5 S_4+4 P_1^2 P_4^3 S_2-12 P_2 P_4^3 S_2-18 P_1^2 P_4 P_5 S_2\cr\cr
&&+54 P_2 P_4 P_5 S_2+18 P_1^2 P_6 S_2-54 P_2 P_6 S_2+4 P_1^3 P_4^2 S_3-18 P_1 P_2 P_4^2 S_3+18 P_3 P_4^2 S_3\cr\cr
&&-12 P_1^3 P_5 S_3+54 P_1 P_2 P_5 S_3-54 P_3 P_5 S_3-47 P_1^2 P_4^2 S_4-99 P_2 P_4^2 S_4+111 P_1^2 P_5 S_4\cr\cr
&&+45 P_2 P_5 S_4+48 P_1 P_4 S_5\Big)\cr\cr
&\equiv&f_{1111}(\{P_i\})
\eea
which are derived in a similar way to the constraints obtained above.

\section{Constraint Structure for $N=3$ bosons in $d=2$ dimensions}\label{confirm}

The partition function derived in Section \ref{NiceQRep} is given by
\bea
Z(t_1,t_2)&=&\frac{N(t_1,t_2)}{(1-t_1)(1-t_1^2)(1-t_1^3)(1-t_2)(1-t_2^2)(1-t_2^3)(1-t_1t_2) (1-t_1^2t_2) (1-t_1t_2^2)}\cr\cr
&&
\eea
with
\bea
N(t_1,t_2)&=&1-t_1^3t_2^2-t_1^4 t_2^2-t_1^2t_2^3-t_1^2t_2^4-t_1^3t_2^3-t_1^4t_2^4\cr\cr &&+t_1^5t_2^3+2t_1^4t_2^4+t_1^5t_2^4+ t_1^3t_2^5+t_1^4t_2^5+t_1^6t_2^5+t_1^5t_2^6\cr\cr
&&-t_1^5t_2^6-t_1^6t_2^5-t_1^7t_2^7
\eea
The complexity of the numerator in the Hilbert series reflects the presence of nontrivial relations among the generators -- specifically, a system of constraints that are themselves interdependent. This intricate structure implies that the space of invariants is not freely generated. To confirm this interpretation, we start by examining the leading negative terms in the Hilbert series. These terms encode the first layer of constraints among the generators. The corresponding constraint polynomials are derived in detail in the appendices; their explicit forms are essential for verifying the identities presented in equations (\ref{relsbetcons}) and (\ref{relsbetrels}) below
\bea
-t_1^3t_2^2&:&\chi_1\,\,=\,\,S_1 S_2-f_{12}(\{P_i\})\,\,=\,\,0\cr\cr
-t_1^4 t_2^2&:& \chi_2\,\,=\,\,(S_2)^2-f_{22}(\{P_i\})\,\,=\,\,0\cr\cr
-t_1^2t_2^3&:&\chi_3\,\,=\,\,S_1 S_3-f_{13}(\{P_i\})\,\,=\,\,0\cr\cr
-t_1^2 t_2^4&:&\chi_4\,\,=\,\,(S_3)^2-f_{33}(\{P_i\})\,\,=\,\,0\cr\cr
-t_1^3t_2^3&:&\chi_5\,\,=\,\,S_2 S_3-f_{23}(\{P_i\})\,\,=\,\,0\cr\cr
-t_1^4t_2^4&:&\chi_6\,\,=\,\,(S_1)^4-f_{1111}(\{P_i\})\,\,=\,\,0
\eea
The next positive terms in $N(t_1,t_2)$ record the relations between the constraints. There are 8 such independent relations
\bea
t_1^5t_2^3&:&\eta_1\,\,=\,\,S_1\chi_2-S_2\chi_1\,\,=\,\,0\cr\cr
t_1^4t_2^4&:&\eta_2\,\,=\,\,S_3\chi_1-S_2\chi_3\,\,=\,\,0\cr\cr
t_1^4t_2^4&:&\eta_3\,\,=\,\,S_3\chi_1-S_1\chi_5\,\,=\,\,0\cr\cr
t_1^5t_2^4&:&\eta_4\,\,=\,\,S_3\chi_2-S_2\chi_5\,\,=\,\,0\cr\cr
t_1^3t_2^5&:&\eta_5\,\,=\,\,S_1\chi_4-S_3\chi_3\,\,=\,\,0\cr\cr
t_1^4t_2^5&:&\eta_6\,\,=\,\,S_2\chi_4-S_3\chi_5\,\,=\,\,0\cr\cr
t_1^6t_2^5&:&\eta_7\,\,=\,\,S_2\chi_6-S_1^3\chi_1\,\,=\,\,0\cr\cr
t_1^5t_2^6&:&\eta_8\,\,=\,\,S_1^3\chi_3-S_3\chi_6\,\,=\,\,0\label{relsbetcons}
\eea
The last three negative terms represent three independent relations between the relations
\bea
-t_1^5t_2^6&:&\phi_1\,\,=\,\,S_1\eta_6+S_3\eta_2-S_2\eta_5\,\,=\,\,0\cr\cr
-t_1^6t_2^5&:&\phi_2\,\,=\,\,S_3\eta_1-S_2\eta_3+S_1\eta_4\,\,=\,\,0\cr\cr
-t_1^7t_2^7&:&\phi_3\,\,=\,\,S_2\eta_8+S_3+S_1^3\eta_2\,\,=\,\,0\label{relsbetrels}
\eea
This analysis demonstrates that the partition function is perfectly consistent with the algebra generated by the primary invariants and the single particle secondary invariants.

\section{Details of the construction for $C_{42}$}\label{AlgExec}

In this appendix we summarize the details of the analysis carried out in \cite{pgeometry}.

Using the equivalence between $A$ and $B$, we can restrict to bidegrees $(r, s)$ with $r \geq s$, and obtain the remaining invariants by swapping $A\leftrightarrow B$. Up to degree $4$ all traces are simply linear combinations of the generators. In bidegree~$(5, 0)$, we find 
$$\Tr(A^5)=\frac{5}{6}\Tr(A^2) \Tr(A^3)=\frac{5}{6}P_{3}P_{6}$$ 
by using the trace relations. In bidegree~$(4, 1)$, again using the trace relations, we find
\bea
\Tr(A^4B) &= \dfrac{1}{2} \Tr(A^2) \Tr(A^2B) + \dfrac{1}{3} \Tr(A^3) \Tr(AB)
= \dfrac{1}{2}P_3 P_7 + \dfrac{1}{3}P_4 P_6\label{forAAAAB}
\eea
Now consider bidegree~$(3, 2)$. There are two single traces operators that are not $CH_4$: $\Tr(A^3B^2)$ and~$\Tr(A^2BAB)$ and there is a single breaking pair:~$(B^2, A^4B)$. The Poisson bracket of the traces of this pair is
\bea
\{\Tr(B^2), \Tr(A^4B)\} = -4 \Tr(A^3B^2) -4 \Tr(A^2BAB) \label{frstbrp}
\eea
On the other hand, by making use of (\ref{forAAAAB}) we find
\bea
\{\Tr(B^2), \Tr(A^4B)\} &=& \{ \Tr(B^2), \dfrac{1}{2} \Tr(A^2) \Tr(A^2B) + \dfrac{1}{3} \Tr(A^3) \Tr(AB) \} \cr\cr
&=& 4 \Tr(AB) \Tr(AB^2) + 2 \Tr(A^2B) \Tr(B^2) + \dfrac{2}{3} \Tr(A^2) \Tr(B^3)\cr
&&\label{scndbrp}
\eea
Equating (\ref{frstbrp}) and (\ref{scndbrp}) we find
\bea
\Tr(A^3B^2)+\Tr(A^2BAB)=-\Tr(AB) \Tr(AB^2)-\frac12 \Tr(A^2B) \Tr(B^2)-\dfrac{2}{12} \Tr(A^2) \Tr(B^3)
\nonumber
\eea
Introducing the generator
\bea
S_1 = \Tr([A, B]^2A) = -\Tr(A^3B^2) + \Tr(A^2BAB),\nonumber
\eea
and solving the last two equations above, we find
\begin{align*}
\Tr(A^3B^2) &= \frac{1}{12} (P_5 P_6+6 P_4 P_7+3 P_3 P_8-6 S_1), \\
\Tr(A^2BAB) &=\frac{1}{12} (P_5 P_6+6 P_4 P_7+3 P_3 P_8+6 S_1).
\end{align*}
By swapping $A\leftrightarrow B$ this determines all single traces with degree $5$.

For degrees up to 11, there are no relations among the trace monomials (necklaces), so every trace has a unique expression in terms of the generators. Proceed inductively by degree $n$, applying the following algorithm:
\begin{itemize}
\item[1.] Apply the Cayley–Hamilton theorem to reduce traces of $CH_4$-type necklaces to expressions involving lower-degree traces.
\item[2.] For non-$CH_4$ necklaces, use breaking pairs to compute Poisson brackets between traces of necklaces.
These brackets give linear combinations of traces, some of which are already known in terms of generators.
\item[3.] Exactly as we did for the example described above, substitute known expressions into the brackets to generate linear equations for the unknown trace terms. The properties of breaking pairs ensures that at least one of $\Tr(w_i)$ can be replaced by an algebraic expression of lower degree terms.
\item[4.] By systematically repeating this (starting from highest bidegree and going down), a linear system is built whose solution gives all trace expressions in terms of the generators.
\end{itemize}
To improve efficiency, it's not necessary to use all breaking pairs—randomly chosen pairs suffice, as long as they generate a fully determined system.

Starting from degree 12, relations among traces (i.e., nontrivial dependencies) begin to appear. As a result, the expression of a trace in terms of generators is no longer unique—only its equivalence class modulo the ideal of relations matters. Key points:
\begin{itemize}
\item[1.] The same algorithm as above is applied, but now the resulting linear system becomes inconsistent due to the emergence of relations.
\item[2.] These inconsistencies are used to identify new relations among the generators, which are then added to an ideal $I$ that grows inductively with each degree.
\item[3.] A special subtlety occurs at bidegree $(\frac{n}{2},\frac{n}{2})$ (for even $n$), where the breaking pair method may miss a single relation.
\item[4.] To find the missing piece, consider $\Tr([A,B]^{n\over 2})$ and evaluate it in two different ways:
\begin{itemize}
\item[$\bullet$] One using direct expansion and Cayley-Hamilton reductions.
\item[$\bullet$] One using already known expressions for lower-degree traces.
\end{itemize}
The difference provides the missing relation, which is then added to the ideal.
\item[5.] After generating all relations at degree $n$, compute the Hilbert series of the ideal $I$ and compare it to the expected one. If they don’t match, the process continues at higher degree.
\end{itemize}

There is one potential point of confusion that we should clarify. How is it possible that relations appear? The Hironaka decomposition implies that the expression of a given necklace in terms of primary and secondary invariants is unique. 

The Hironaka decomposition says that the ring of invariants $\mathbb{C}[A,B]^{U(N)}$ has the structure:
\bea
\mathbb{C}[A,B]^{U(N)}=\mathbb{C}[P_1,P_2,\cdots,P_{1+(d-1)N^2}]\cdot{\rm Span}_{\mathbb{C}}\{S_0,\cdots,S_{K-1}\}
\eea
where $K$ is the total number of secondary invariants. This decomposition guarantees that every invariant has a unique expression as a linear combination of the secondary invariants, with coefficients that are polynomials in the primary invariants.

Why then do relations appear? We are starting from a redundant basis (all single trace operators), and writing them in terms of a minimal basis (the primary and secondary invariants). This process involves solving linear systems of equations -- but starting from degree 12, not all traces of necklaces are independent any more. That is, there are linear relations among traces, coming from trace identities. These relations must be modded out to identify the correct linear combinations that span the space of invariants. The relations that ``appear" in degree $\ge 12$ are relations among traces of necklaces -- they reflect linear dependencies among these traces, not a failure of uniqueness in the Hironaka decomposition.

The Hironaka decomposition gives unique expressions for invariants in terms of primary and secondary invariants. The relations that appear are among single traces, and are used to identify which combinations of traces form the secondary invariants. So these ``relations" are the mechanism by which the uniqueness guaranteed by the Hironaka structure is enforced.

\section{Hilbert Series and the Hironaka Decomposition}

In this Appendix we will argue, with a simple example, that to demonstrate that an algebra admits a Hironaka decomposition it is not good enough to demonstrate that the Hilbert series takes the form (\ref{HironakaForm}). Our example is given by the ring
\bea
R&=&\mathbb{C}[x,y]/\langle xy\rangle
\eea
This is the ring of polynomials in $x$, $y$, but with any product of $x$'s and $y$'s set to zero. Thus a general member of this ring is given by the sum of two single variable polynomials
\bea
f(x,y)&=&g_1(x)+g_2(y)\in R
\eea
Since $x$ and $y$ are algebraically independent, $R$ has Krull dimension 2. It has depth zero because even the condition
\bea
x f(x,y)=0
\eea
does not imply that $f(x,y)=0$. Since Krull dimension does not equal depth, this ring is not Cohen-Macaulay. The Hilbert series of $R$ is easily computed as
\bea
H_R(t)&=&{1\over 1-t}+{1\over 1-t}-1\,\,=\,\,\frac{1+t}{1-t}
\eea
which is of the form (\ref{HironakaForm}), even though $R$ does not admit a Hironaka decomposition.

\end{appendix}

\end{document}